\documentclass[pdflatex,sn-mathphys-num]{sn-jnl}

\usepackage{graphicx}%
\usepackage{multirow}%
\usepackage{amsmath,amssymb,amsfonts}%
\usepackage{amsthm}%
\usepackage{mathrsfs}%
\usepackage[title]{appendix}%
\usepackage{xcolor}%
\usepackage{textcomp}%
\usepackage{manyfoot}%
\usepackage{booktabs}%
\usepackage{algorithm}%
\usepackage{algorithmicx}%
\usepackage{algpseudocode}%
\usepackage{listings}%
\usepackage{doi}
\usepackage{subcaption}

\theoremstyle{thmstyleone}%
\theoremstyle{thmstyletwo}%

\theoremstyle{thmstylethree}%

\begin{document}

\title[Article Title]{A novel approach for quantum financial simulation and quantum state preparation}


\author*[1,2]{\fnm{Yen-Jui} \sur{Chang}}\email{d09222012@ntu.edu.tw}
\author[1,2]{\fnm{Wei-Ting} \sur{Wang}}\email{aroe01325@gmail.com}
\author[4]{\fnm{Hao-Yuan} \sur{Chen}}\email{hc118@student.london.ac.uk}
\author[3]{\fnm{Shih-Wei} \sur{Liao}}\email{liao@csie.ntu.edu.tw}
\author[1,5]{\fnm{Ching-Ray} \sur{Chang}}\email{crchang@phys.ntu.edu.tw}

\affil*[1]{\orgdiv{Department of Physics}, \orgname{National Taiwan University}, \orgaddress{\street{No. 1, Sec. 4, Roosevelt Rd.}, \city{Taipei}, \postcode{106315}, \country{Taiwan}}}
\affil*[2]{\orgname{NTU-IBM Quantum Hub}, \orgaddress{\street{No. 1, Sec. 4, Roosevelt Rd.}, \city{Taipei}, \postcode{106315}, \country{Taiwan}}}

\affil[3]{\orgdiv{Department of Computer Science and Information Engineering}, \orgname{National Taiwan University}, \orgaddress{\street{No. 1, Sec. 4, Roosevelt Rd.}, \city{Taipei}, \postcode{106315}, \country{Taiwan}}}

\affil[4]{\orgdiv{Department of Computer Science}, \orgname{University of London}, \orgaddress{\street{Senate House
Malet Street London WC1E 7HU}, \city{London},  \country{United Kingdom}}}

\affil[5]{\orgdiv{Quantum Information Center}, \orgname{Chung Yuan Christian University}, \orgaddress{\street{No. 200, Zhongbei Rd.}, \city{Taoyuan City}, \postcode{320314}, \country{Taiwan}}}

\abstract{Quantum state preparation is crucial in quantum computing for various applications, notably in quantum simulation where a quantum state must represent the simulated system. This study unveils the multi-Split-Steps Quantum Walk (multi-SSQW), an advanced simulation algorithm using parameterized quantum circuits (PQC) and a variational solver to manage complex probability distributions. Enhanced from the traditional split-steps quantum walk (SSQW) to include multi-agent decision-making, the multi-SSQW is adept at financial market modeling. It leverages quantum computation to accurately model intricate financial distributions and scenarios, offering key insights for financial analysis and strategic decisions. The algorithm's flexibility, reliable convergence, and rapid computation make it a powerful tool for fast-paced financial market predictions.}




\maketitle

\section{Introduction}\label{sec1}

Since the 1950s, digital computing has significantly advanced scientific capability to address complex issues quantitatively, laying the groundwork for modern quantitative finance and transforming financial analysis and decision-making. Quantum mechanics, with its principles of superposition, entanglement, and interference, offers groundbreaking solutions to challenges beyond classical computing's reach. The fusion of computer science and quantum physics has unveiled potential for addressing problems that classical computing finds intractable. Quantum algorithms, leveraging the inherent uncertainty of quantum mechanics, present innovative ways to model complex probability distributions, providing a fresh approach to computational methods and enhancing operations like factorization\cite{Shor1994} and simulation\cite{Feynman1982, Trabesinger2012} more efficiently than classical computers.

Uncertainty is an inherent nature of financial markets, manifesting itself in the unpredictability of the cognitive processes of market participants, their decision-making routines, and the general macroeconomic conditions and structural dynamics of the financial market, which directly influence asset valuation. This concept is strikingly similar to quantum theory. Quantum computing is particularly promising for the financial sector, which stands to benefit significantly from this innovation. This is primarily due to many financial scenarios that could be resolved using quantum algorithms. Our ambitions focus on developing quantum states that encapsulate the inherent uncertainties that pervade financial markets, which can be processed using quantum computers. Additionally, we aim to devise quantum algorithms to simulate the dynamic nature of financial systems.

In traditional finance, the random walk theory is a prevalent model. This theory, first postulated by the French mathematician Louis Bachelier\cite{Bachelier:RW}, posits that the trajectory of stock prices is essentially random. In other words, the future price of a stock is independent of its past prices, making it impossible to predict a stock's future trajectory based on historical data alone. This idea forms the basis of the Efficient Market Hypothesis\cite{Fama_EMH}, stating that all available information is already incorporated into the stock's current price and changes to that price will only be triggered by unforeseen events. Stock prices are the product of an ongoing interplay of buying and selling transactions from all participants in the stock market. This dynamic process reflects the collective sentiment, beliefs, and actions of all these market participants. Therefore, the random walk theory of stock prices doesn't mean prices are entirely chaotic, but rather that they evolve based on the aggregate of numerous decisions made by market participants, often in response to new information.

Given that uncertainty is a fundamental characteristic shared by both finance and quantum mechanics, it is promising to employ quantum principles for the simulation of financial markets. 
The potential applications of quantum computing in the financial sector\cite{9222275,Herman2022} are incredibly vast. More recent work has focused on the quantum algorithm for amplitude estimation\cite{Brassard_2002} and Monte Carlo with the pricing of financial derivatives\cite{Rebentrost2018, Zoufal:qGAN2019,Stamatopoulos_2020,Stamatopoulos_2022,Dong_An_2021,Chakrabarti_2021}. Ref. \cite{Brassard_2002}, which builds upon Grover's quantum search method to improve the likelihood of identifying desired outcomes in quantum algorithms without needing to know the success probabilities in advance. Ref. \cite{Rebentrost2018} presents a quantum algorithm for Monte Carlo pricing of financial derivatives, demonstrating how quantum superposition and circuits can implement payoff functions and extract prices through quantum measurements. Ref. \cite{Stamatopoulos_2020} details a method for option pricing using quantum computing. This method leverages amplitude estimation to achieve a quadratic speed increase over traditional Monte Carlo methods, showcasing significant advancements in quantum algorithm applications and financial modeling techniques. Furthermore, Ref.\cite{Zoufal:qGAN2019,Stamatopoulos_2022,Dong_An_2021,Chakrabarti_2021} explore various aspects, from implementing quantum computational finance and option pricing to leveraging quantum advantage in market risk assessment and stochastic differential equations. Each study contributes to the broader understanding of how quantum algorithms can offer a more efficient, accurate and comprehensive approach to financial simulations, surpassing traditional computational methods and providing new insights into quantum finance's potential.

In leveraging the principles of quantum mechanics for financial market analysis, we adopt a novel approach by interpreting states of financial uncertainty through the lens of quantum states. This perspective allows us to model potential financial outcomes within the quantum framework, thus providing a more nuanced understanding of market dynamics.  Equation \ref{Eq:eq1}

\begin{equation}
|\psi\rangle = \sum_{i} \sqrt{p_{i}} \;|i\rangle 
\label{Eq:eq1}
\end{equation}
illustrates the concept of superposition in quantum mechanics, where \(|\psi\rangle\) represents the superposition state of the system. Each potential outcome in the financial market is analogized to a quantum state \(|i\rangle\), with \(p_{i}\) denoting the probability of the system being in state \(|i\rangle\). This superposition principle enables the encapsulation of multiple possible financial outcomes within a single quantum state, offering a powerful tool for modeling financial uncertainty and making predictions based on quantum probabilistic outcomes.

In simulating the evolution of financial markets through a quantum mechanics framework, we employ a model that mirrors the evolution of quantum systems. Equation \ref{Eq:eq2} captures this:

\begin{equation}
U(t)\;|\psi(0)\rangle = |\psi(t)\rangle,
\label{Eq:eq2}
\end{equation}
where \(U(t)\) represents a unitary operator that governs the evolution of the quantum state over time, transitioning the initial state \(|\psi(0)\rangle\) to its future state \(|\psi(t)\rangle\) at time \(t\). In financial markets, the Eq. \ref{Eq:eq2} symbolizes applying quantum evolution principles to model how market states evolve under the influence of various factors, including investor sentiment. To achieve this, we adopt Quantum walks(QW) algorithm as a critical methodology for simulating the nuances of price formation processes influenced by investor behavior. Unlike traditional models, QW introduces complexity and precision by leveraging quantum systems' inherent probabilistic nature and superposition capabilities. The quantum evolution model represented by Eq. \ref{Eq:eq2} opens new avenues for understanding and predicting financial market behaviors more comprehensively and nuancedly.

Quantum walks, the quantum mechanical counterparts to classical random walks, offer an innovative financial simulation approach. Whereas classical random walks allow a system to move randomly from one point to another, quantum walks incorporate the principles of superposition and entanglement, enabling a quantum system to exist in multiple states simultaneously and to explore many paths at once.In a financial context, a quantum walk could simulate multiple behaviors in parallel, potentially capturing the complexity and randomness of financial markets more accurately than a classical random walk. Therefore, applying the quantum walk algorithm to financial simulations could provide a more detailed and comprehensive understanding of potential outcomes and market behavior. This extension of random walks to the quantum realm could open new financial modeling and risk management.

Using the quantum walk algorithm in the financial simulations allows us to set up the quantum state in a superposition of states, each corresponding to a potential financial outcome. This method generates designated amplitudes for these states in a way that closely resembles a targeted probability distribution and could also provide a novel method for quantum state preparation. Quantum state preparation is a crucial step in quantum computing and quantum simulations. The ability to prepare quantum states allows for parallel information processing, enabling quantum computers to solve specific quantum algorithms\cite{Brassard_2002,HHL:2009,PCA:2014,QSVM:2019,QC2021,Montanaro2016} significantly faster than classical computers. The referenced works \cite{HHL:2009,PCA:2014,QSVM:2019} highlight significant advancements in quantum computing applications. Harrow, Hassidim, and Lloyd \cite{HHL:2009} introduce a quantum algorithm for solving linear systems of equations, offering a potential exponential speedup over classical methods. Lloyd, Mohseni, and Rebentrost \cite{PCA:2014} propose a quantum principal component analysis, paving the way for more efficient data processing. Havlíček et al. \cite{QSVM:2019} discuss the application of quantum-enhanced feature spaces in supervised learning, suggesting a new approach to machine learning with quantum computing. These studies collectively demonstrate the broad potential of quantum algorithms in solving complex computational problems across various domains.
Furthermore, Ref. \cite{QC2021,Montanaro2016} emphasize the transformative potential of quantum computing, illustrating key advancements in quantum algorithms and their wide-ranging applications. These works offer insight into quantum computing's ability to tackle complex problems more efficiently than conventional approaches, spotlighting the dynamic progress and future possibilities in various sectors through quantum solutions.

\begin{figure}[htb]
\centering
\includegraphics[width=10cm]{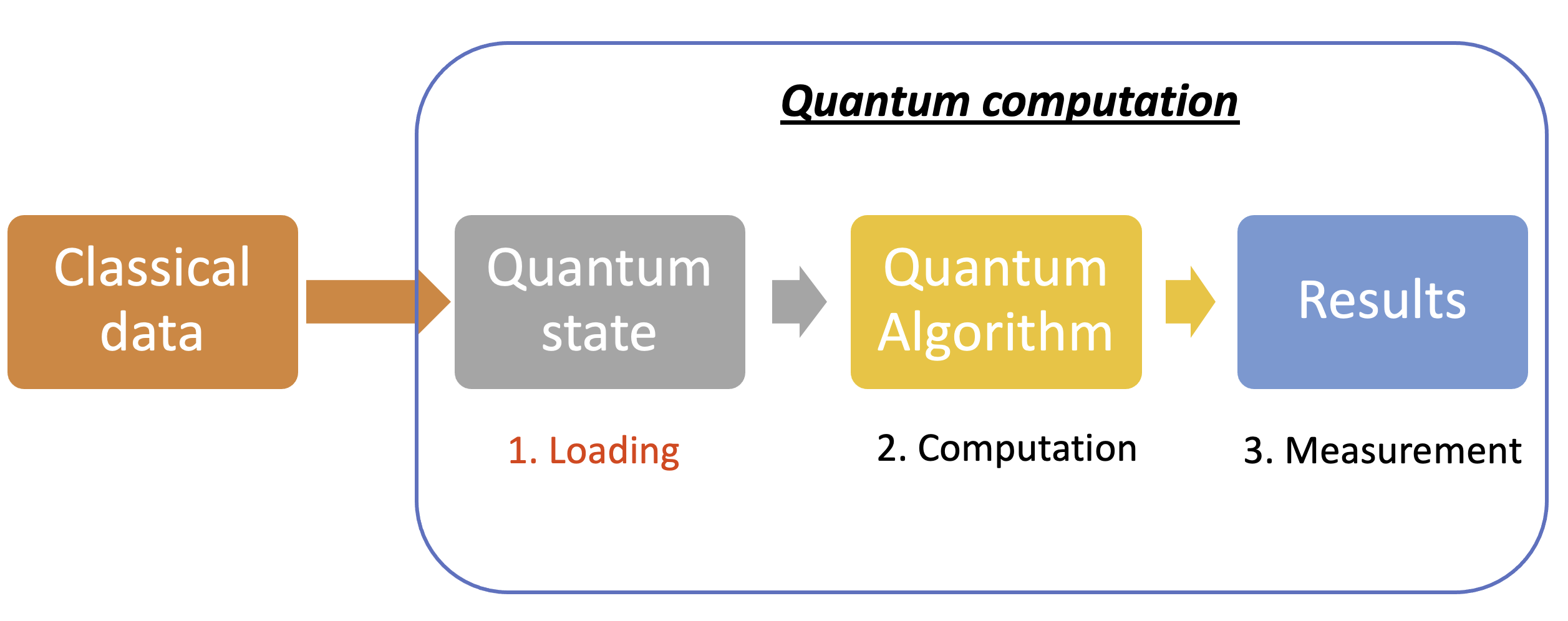}
\caption{ The three generic steps to realize the quantum advantage. }
\label{fig:FIG1}
\end{figure}

The achievement of quantum advantage is considered an essential goal in quantum computing. The three steps of realizing quantum advantage: quantum state preparation, quantum algorithm computation, and result measurements, are shown in Fig.\ref{fig:FIG1}. 
The ability to prepare quantum states in quantum circuits has many conceivable applications. Efficient methods to simulate probability distributions in quantum circuits are critical, as an ineffective and inefficient quantum state preparation algorithm could diminish the potential impact of quantum computing\cite{PhysRevLett.122.020502}. The randomness of quantum systems has been demonstrated \cite{Choi2023} with the distribution of states close to the Haar-random. Haar-random\cite{Nielsen2000,Guise2018} is significant in quantum computations, as they can be used to generate arbitrary quantum states and perform quantum gates that enable applications of these quantum devices in a much broader context. These generic quantum state preparation methods have been designed\cite{Zoufal:qGAN2019,Grover:loading2002,Rocchetto2018QSP, Kalayn:Loading2022, Parl2021QSP,Zhang2022QSP,Yuan2023QSP}. Grover has proposed a scheme \cite{Grover:loading2002} to generate the probability distributions. It shows how to generate a superposition of quantum states by taking an ancilla register that performs a controlled rotation of angle $\theta_{i}$:
\begin{equation} \label{eq:controlratation}
\sqrt{{p_{i}}} \;|i\rangle \to \sqrt{{p_{i}}} \;|i\rangle \otimes (\cos \theta\;|0\rangle + \sin \theta |1\rangle).
\end{equation}

Using ancilla qubits makes it possible to reduce the circuit depth, resulting in complexity mitigation at a scale less than exponential\cite{Zhang2022QSP}.

Moreover, Rocchetto has proposed a method\cite{Rocchetto2018QSP} based on variational autoencoders to encode the probability distribution of quantum states and benchmark the performance of deep networks in states. Quantum Generative Adversarial Networks (qGAN) have been demonstrated to load distributions\cite{Zoufal:qGAN2019}. The qGAN combines a quantum generator and a classical discriminator to learn the probability distribution of classical training data. The quantum generator, a parametrized quantum channel, is trained to convert an input state of n-qubits, represented by $\;|\psi\rangle$, into an output state of n-qubits. A method using variational solvers to fix gate rotation parameters has been proposed to generate symmetrical and asymmetric probability distributions\cite{Kalayn:Loading2022}. The authors demonstrated trajectories of the individual quantum states to understand the effect of an ancilla register to control rotation.  

In this study, we introduce a novel methodology based on Eq.(2) that incorporates multi-SSQW, and expanded the concept of Single-Split-Step Quantum Walk (SSQW), into the simulation of the financial system and quantum state preparation. Multi-SSQW leverages an ancilla qubit as a coin space to control the position space, representing the targeted quantum state. The structure of the paper unfolds as follows:

Firstly, we rethink the theoretical underpinnings of quantum walks and shed light on their role in simulating financial pricing and preparing quantum states. Next, we engineer the methodology of multi-SSQW and apply the approach to a range of test cases, on financial pricing process simulating and quantum state preparation, including normal, log-normal, and binomial distributions, with a quantum simulator accessible via IBM Q Experience. We then demonstrate using the multi-SSQW method to facilitate quantum benefit in financial derivative pricing. 

\section{Multi-split-steps quantum walk (multi-SSQW)}\label{sec2}
\subsection{Mathematical Model}
The multi-SSQW represents an expansion of conventional quantum walks, extending its capabilities and potential applications. QW\cite{Feynman:quantumcomputers,Aharonov1993,Childs:quantumwalk,Mallick:DCA2016,Rajauria:DTQW,Venegas_Andraca_2012,10.1145/780542.780552,Childs2004} are used as a foundation for generating models of controlled quantum simulation. 
The evolution of QW marks a crucial chapter in quantum computing, originating from Feynman's\cite{Feynman:quantumcomputers} early ideas on quantum mechanics for computational use. Over time, the scope of research has expanded to include quantum random walks\cite{Aharonov1993,Venegas_Andraca_2012}, their role in universal computation, and their contribution to improving algorithmic efficiency and enabling quantum simulations. Significant strides in QW utilization for quantum computing have been marked by Childs' work on universal computation\cite{Childs:quantumwalk,Childs2004,10.1145/780542.780552}, the integration of Dirac cellular automata\cite{Mallick:DCA2016}, and the innovative application of machine learning to fine-tune QW parameters\cite{Rajauria:DTQW}. These developments underscore the growing capability of QW to facilitate complex quantum computing operations. These contributions have deepened our comprehension of QW, from their theoretical underpinnings to practical applications in enhancing computational processes, showcasing a dynamic field of study.
QW lays the groundwork for controlled quantum simulations. They offer a versatile framework for mimicking quantum-mechanical behaviors by adjusting parameters and coin operators within the walks. QW is broadly categorized into discrete-time (DTQW) and continuous-time (CTQW) models, each with distinct characteristics that render them effective for specific quantum computing operations.

QW enables the walker to simulate several quantum-mechanical phenomena by tuning a QW's parameters and evolution coin operators. Here, we will focus only on the one-dimensional DTQW. A classical walk can be described using just a position Hilbert space, while a DTQW requires an additional coin Hilbert space to express its dynamics fully. This coin space represents the internal state of the walker and is necessary to capture the controlled dynamics of the walker.
Hilbert space of QW is defined as follows. 
\begin{equation} \label{eq:HilbertSpace}
\mathcal{H} = \mathcal{H}_{c}  \otimes \mathcal{H}_{p},
\end{equation}
where $\mathcal{H}_{c}$ is the coin Hilbert space and  $\mathcal{H}_{p}$ is the position Hilbert space. The coin Hilbert space for one-dimensional DTQW has the basis states $\{\;|\uparrow\rangle = \begin{pmatrix}
  1\\ 
  0
\end{pmatrix} , \;|\downarrow\rangle = \begin{pmatrix}
  0\\ 
  1
\end{pmatrix}\}$ and the position Hilbert space is defined by the basis states $\;|x\rangle $ where $x\in Z$. The probability amplitude of the quantum state at position $x$ can be represented by
\begin{equation} \label{eq:amplitudeofthequantumstate}
\Psi(x,t) \rangle =
\begin{pmatrix} 
\Psi^{\uparrow} (x,t) \\ \Psi^{\downarrow} (x,t) 
\end{pmatrix},	
\end{equation}
where describe the state of DTQW with two internal degrees of freedom $\{\;|\uparrow\rangle , \;|\downarrow\rangle \}$.
In DTQW, the system's evolution is governed by two unitary operators: the coin operator and the shift operator. The shift operator moves the walker in a superposition of position states, while the coin operator acts on the coin Hilbert space and determines the amplitudes of the position space. The coin Hilbert space represents the internal state of the walker and plays a crucial role in determining the overall dynamics of the system. A universal operator is defined as
\begin{equation} \label{eq:Coinoperator}
   \hat{C} (\theta, \phi ,\lambda) =
   \begin{pmatrix} 
   \cos(\frac{\theta}{2}) & -e^{i \lambda}\sin(\frac{\theta}{2}) \\ e^{i \phi}\sin(\frac{\theta}{2}) &  e^{i(\lambda+\phi)}\cos(\frac{\theta}{2})
   \end{pmatrix},
\end{equation}
where are the three independent parameters and the most general unitary coin operator. Therefore, accurately estimating the coin parameters is essential for effectively using quantum walks as a quantum simulation tool and for further research on modeling realistic dynamics. Finding patterns in complex data can be challenging, but an algorithm that automates the learning process can solve this problem.

The shift operator is an essential part of DTQW. A unitary operator moves the quantum walker in a superposition of position states. The shift operation is defined as,
\begin{equation} \label{eq:shitfoperator}
  \hat{S} = |\downarrow\rangle \langle\downarrow|  \otimes \sum_{x} |x-1\rangle \langle x| + |\uparrow\rangle \langle \uparrow|  \otimes \sum_{x} |x+1\rangle \langle x| .
\end{equation}
In other words, the shift operator shifts the position of the particle one step to the right if the internal state is $|\uparrow\rangle$ or one step to the left $|\downarrow\rangle$.
The initial state of the system is a superposition of the position states, with the internal state of the particle determined by the coin operator and defined as 
\begin{equation} \label{eq:initialstate}
   |\Psi_{0} \rangle = (\alpha|\uparrow \rangle + \beta|\downarrow \rangle) \otimes |x=0 \rangle . 	
\end{equation}
At each time step, the shift operator is applied to the position state after the coin operator is applied to the internal state of the particle. Together, these two operators form the evolution operator of DTQW, which describes the overall dynamics of the system. This process is repeated several times, and the system's final state is a superposition of position states, with the amplitudes determined by the coin operator.
\begin{equation} \label{eq:QWevole}
  \Psi(x,t)=[\hat{S}(\hat{I} \otimes \hat{C})]^{t} |\Psi_{0} \rangle = \hat{W}^{t} |\Psi_{0} \rangle,
\end{equation}
where $\hat{I} =  \sum_{x} |x\rangle \langle x| $ is an identity in space.

\begin{figure}[htb]
\centering
\subcaptionbox{}{\includegraphics[width=4cm]{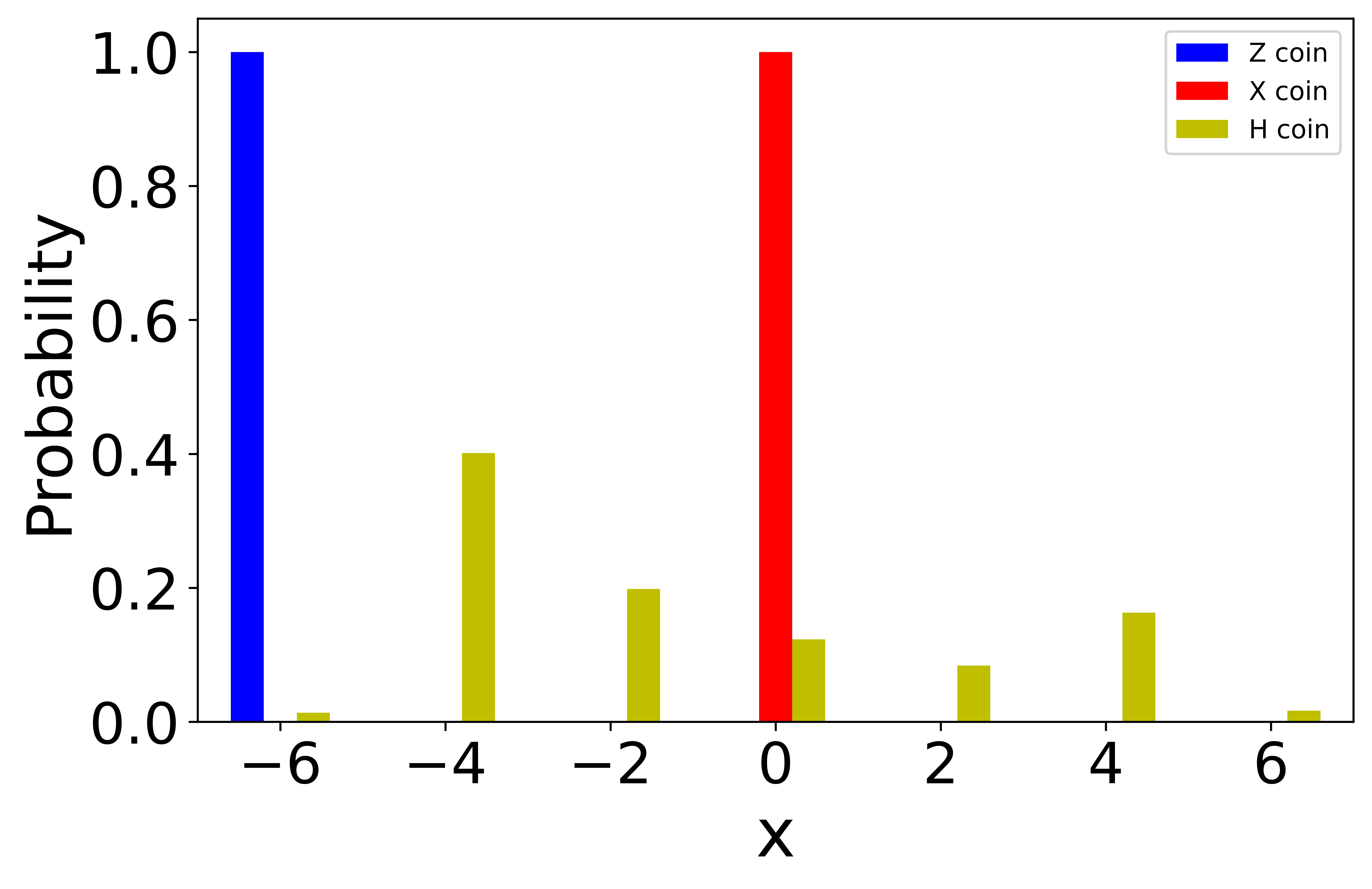}}
\hfill
\subcaptionbox{}{\includegraphics[width=4cm]{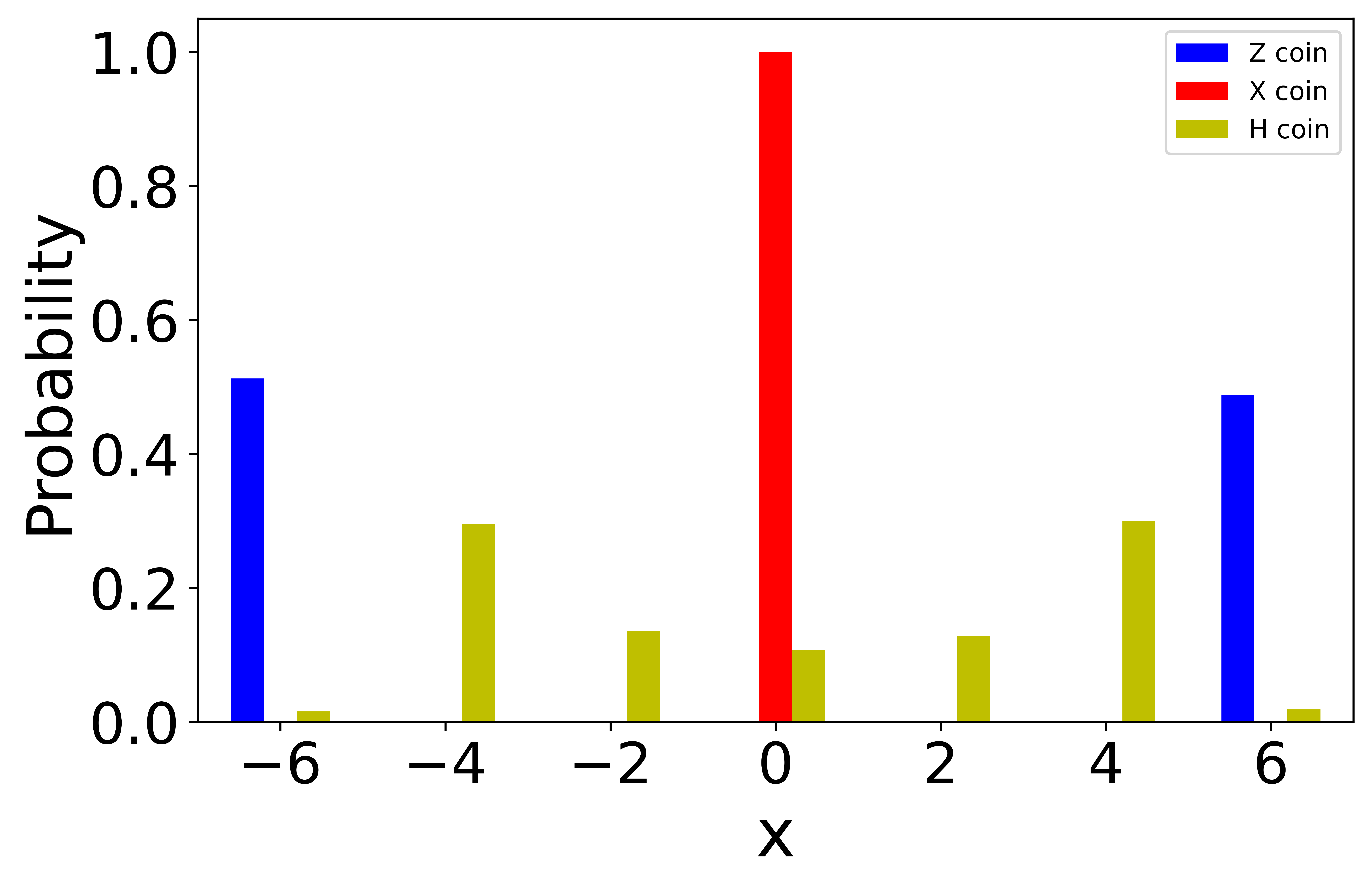}}
\hfill
\subcaptionbox{}{\includegraphics[width=4cm]{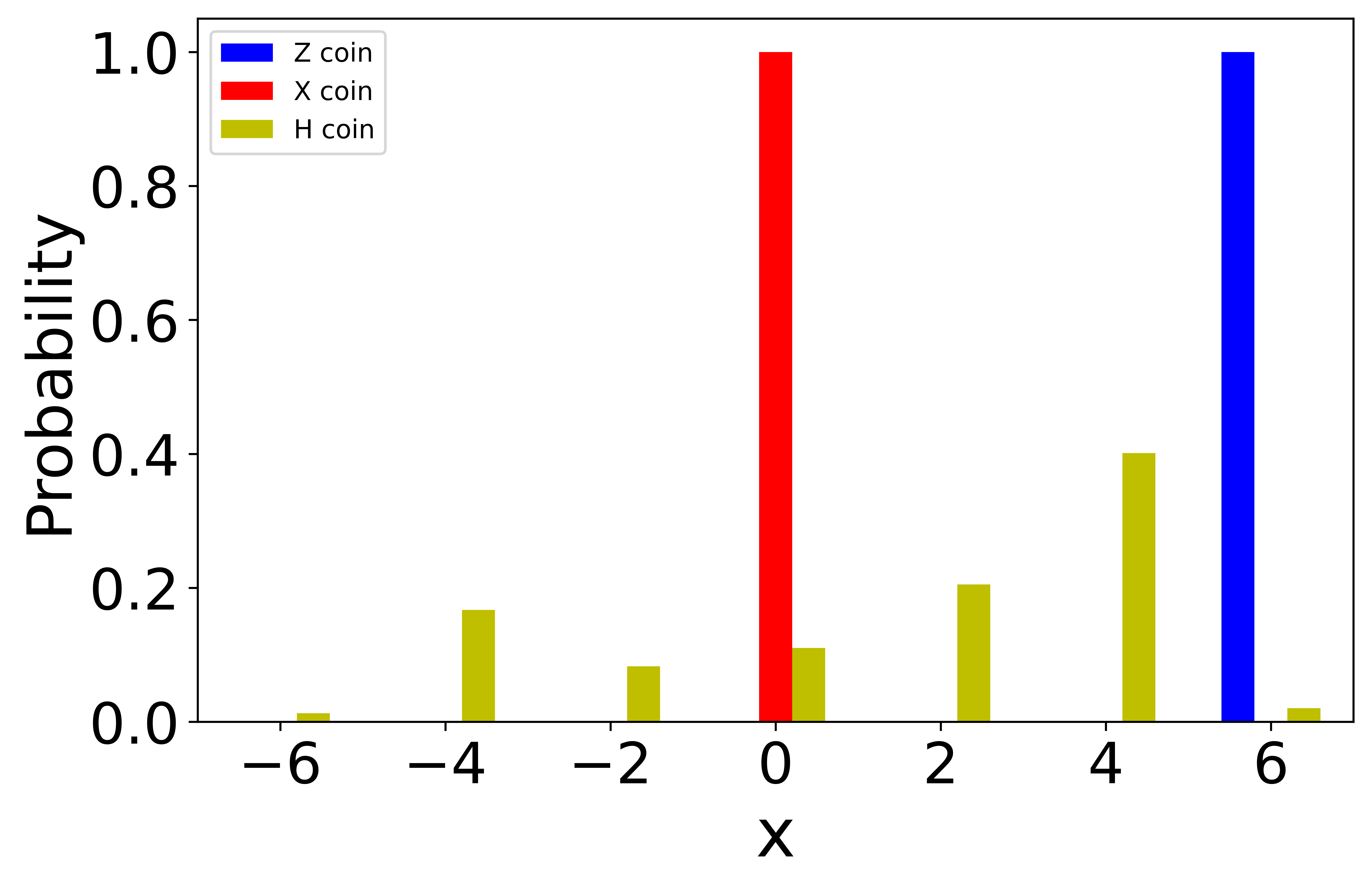}}
\caption{ Demonstrating probability distributions in position space using DTQW with Z, X, and H gates as coin operators and various initial states (a) Initial state $= |0 \rangle$ (b) Initial state $= \frac{1}{\sqrt{2}}(|0 \rangle+i |1 \rangle )$(c) Initial state $= |1 \rangle$ }
\label{fig:FIG2}
\end{figure}

The probability distributions in position space of DTQW \cite{Mallick:DCA2016,Rajauria:DTQW,Venegas_Andraca_2012}, shown in Fig.(\ref{fig:FIG2}), do not resemble the probability distributions in everyday life. In the marketplace, prices are typically determined by the interaction of buyers and sellers. The price of a good or service in the market is established through the agreement between investors. Sentiment-induced buying and selling is an important determinant of stock price variation. The shaping of short-term financial market prices predominantly hinges on the sentiment of investor\cite{Oxenfeldt1973, Ahmed2020}, broadly classified into optimism and pessimism. Investors who are optimistic play a proactive role in investing, which creates an upward push for the stock price. On the contrary, pessimistic investors, who decide to sell and withdraw from the market, generate a downward pull on the stock price. Inspired by the free market economy, we introduce the split-step quantum walk(SSQW)\cite{Matsuzawa_2020,Narimatsu:SSQW} that can be regarded as a financial simulation.

SSQW is a specific type of quantum walk that divides the evolution of the quantum system into two steps, one to the right and one to the left. The evolution operator $\hat{W}$ is divided by a composition of two half-steps,

\begin{equation} \label{eq:SSQW_W}
  \hat{W} = \hat{S}_{-}\hat{C}_{\theta_{2}} \hat{S}_{+}\hat{C}_{\theta_{1}}. 
\end{equation}
where  $\hat{C}_{\theta_{k}}$ is a universal coin operator as Eq.(\ref{eq:Coinoperator}) and shift operators $\hat{S}_{\pm}$ are defined as,
\begin{equation} \label{eq:SSQW_Shift}
\begin{aligned}
   \hat{S}_{+} &= \sum_{x}[|\uparrow \rangle \langle \uparrow| \otimes |x+1\rangle \langle x| +|\downarrow \rangle \langle \downarrow| \otimes |x\rangle \langle x|  ]  \\
   \hat{S}_{-} &= \sum_{x}[|\uparrow \rangle \langle \uparrow| \otimes |x\rangle \langle x| +|\downarrow \rangle \langle \downarrow| \otimes |x-1\rangle \langle x|  ],
\end{aligned}
\end{equation}
where $\hat{S}_{+}$ represents the walker goes to right($|\uparrow \rangle$) or stop in place($|\downarrow \rangle$) and $\hat{S}_{-}$ represents the walker goes to left($|\downarrow\rangle$) or stop in place($|\uparrow \rangle$).

\begin{figure}[htb]
\centering
\subcaptionbox{}{\includegraphics[width=5.5cm]{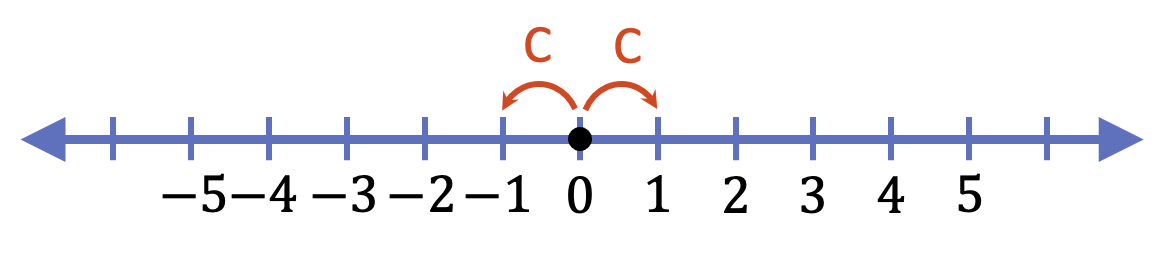}}
\hfill
\subcaptionbox{}{\includegraphics[width=5.5cm]{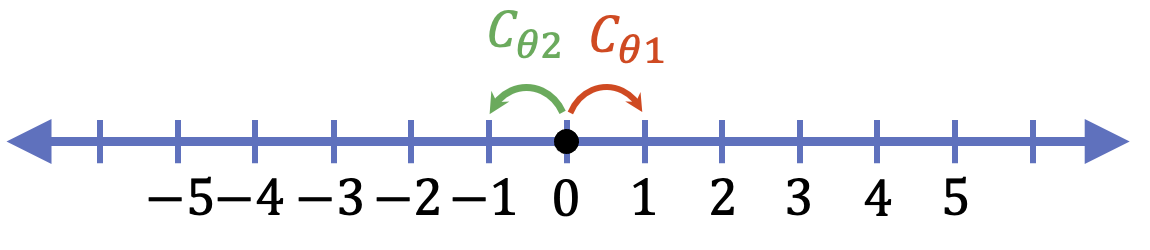}}
\caption{ The scheme of (a)DTQW and (b)SSQW }
\label{fig:FIG3}
\end{figure}
In Fig.\ref{fig:FIG3}, the diagrams depict two types of QW, which are central concepts in quantum computing. Figure (a) illustrates DTQW, where a particle on a line can move to adjacent positions at discrete time intervals, guided by a quantum coin operation $C$ determining its direction. Figure (b) shows SSQW. In this variant, the coin operation is split into two steps, $\hat{C}_{\theta_{1}}$and $\hat{C}_{\theta_{2}}$, allowing for more control over the particle's movement and hence enabling a richer simulation of quantum systems. These models provide frameworks for algorithmic development in quantum simulations and computational tasks, capturing the complexities of quantum mechanics in a controllable process.

In the context of SSQW applied to financial markets, the coin operators, denoted as $\hat{C}_{\theta_{1}}$and $\hat{C}_{\theta_{2}}$, act as quantum analogs to reflect investor sentiment. They control the probability amplitudes for different market states, effectively modeling the decision-making process of investors based on their market sentiments. This mechanism allows for the simulation of market dynamics in a quantum computational framework, offering a nuanced approach to understanding financial fluctuations. $\hat{C}_{\theta_{1}}$ embodies the amplitude of the investor's optimism, influencing the likelihood of a price increase. Conversely, $\hat{C}_{\theta_{2}}$ reflects the amplitude of the investor's pessimism, regulating the probability of a price decrease. This process models the decision-making process of an investor‘s behavior: to buy or not, or to sell or not. This dual-step approach effectively simulates the complex dynamics of trading behavior and investor sentiment. Therefore, investor sentiment is an integral aspect of financial market analysis, much like the concept of superposition in quantum mechanics. With various sentiments and outlooks overlapping and coexisting, it affects not only the trading behavior but also shapes the overall market's perception of risk and value.

The financial market is marked by the participation of a diverse range of investors, each with their unique attitudes and investment strategies. In order to capture this diversity, we build upon existing concepts and introduce the multi-SSQW that corresponds to a multitude of investors. This approach allows us to design a quantum model that simulates various investor sentiments. Such a quantum model aids in simulating trader sentiment within the financial market, enabling us to predict the distribution of short-term financial prices. The scheme for simulating the distribution of short-term financial prices is introduced with a practical approach to find patterns in complex data and map them to a multi-SSQW dynamic system and the architecture has shown below in Fig.(\ref{fig:FIG4}). In the multi-SSQW framework, the U3 gate sets the initial quantum state, encoding the market's overall sentiment. Subsequent unitary operations, represented by two U3 gates for each investor, model individual investor biases towards specific assets. The evolution operator, \( \hat{W} \), is decomposed into two halves—first moving with \( \hat{C}_{\theta_{1}} \) and incrementing (\( \hat{S}_{+} \)) and then applying \( \hat{C}_{\theta_{2}} \) before decrementing (\( \hat{S}_{-} \)). These coin operators, \( \hat{C}_{\theta_{k}} \), and the shift operators, \( \hat{S}_{\pm} \), collectively govern the walker's direction and position, with each \( \hat{W} \) operator reflecting the nuances of investor sentiment towards a financial asset within a quantum simulation model. Figure \ref{fig:FIG4} illustrates the mechanics of a quantum system designed to simulate and analyze financial markets, capturing the dynamics of investor behavior and market fluctuations.

We have extended the concept of SSQW to multi-SSQW, employing multiple walkers to represent investors with diverse investment strategies in the market. In modeling the intrinsic uncertainty in financial markets, we showcase the efficacy of our purpose-built multi-SSQW quantum algorithm and circuitry through its application in replicating the price distribution in real-world stock markets.

\begin{figure}[htb]
\centering
\subcaptionbox{}{\includegraphics[width=12cm]{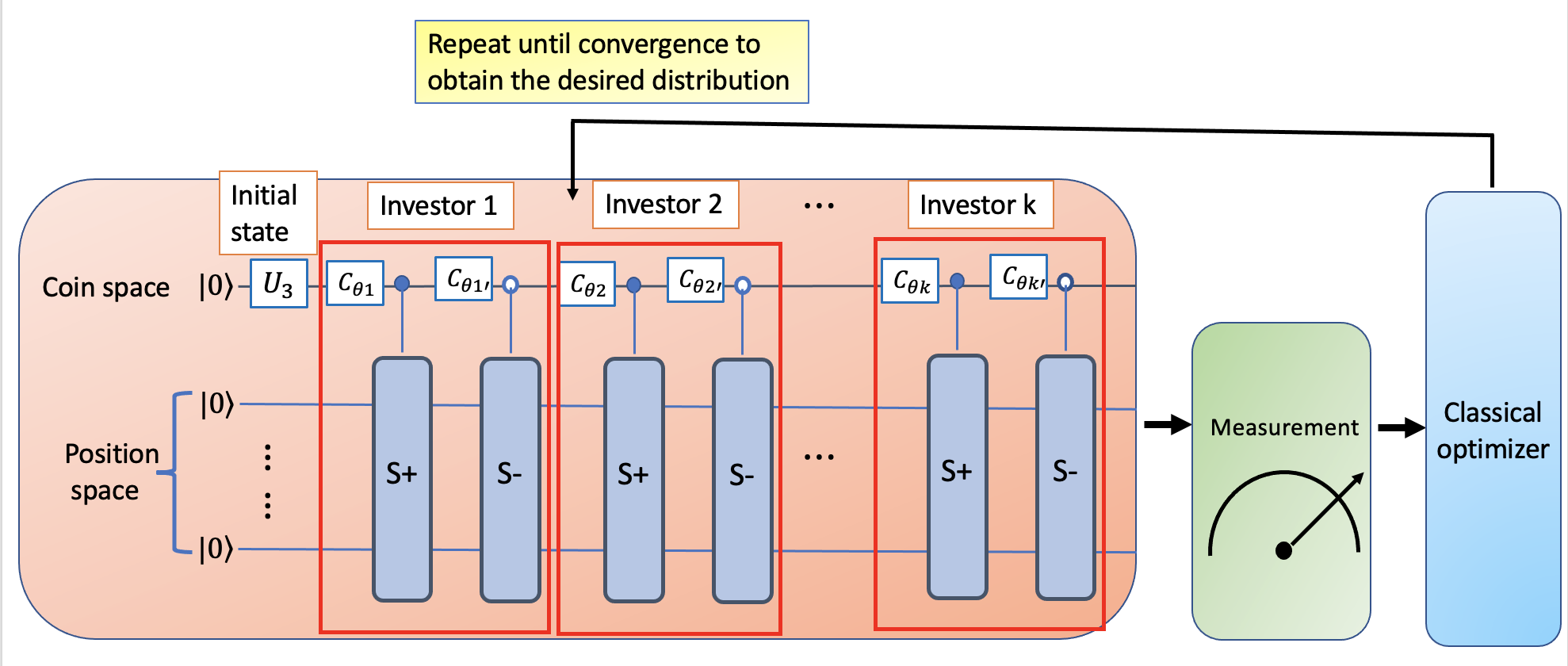}}
\hfill
\subcaptionbox{}{\includegraphics[width=5cm]{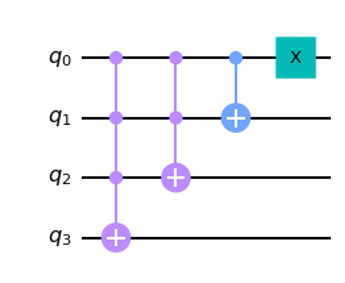}}
\subcaptionbox{}{\includegraphics[width=5cm]{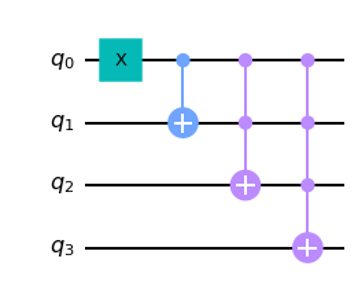}}
\caption{ (a) Illustrates the multi-Split-Steps Quantum Walk (multi-SSQW) setup, showing the initial state and the sequence of operations modeling investors' decision-making. The quantum state evolves iteratively to achieve a targeted distribution, analyzed by a classical optimizer. (b) Depicts the quantum circuit of \( \hat{S}_{+} \), controlling state incrementation. (c) Shows the quantum circuit of \( \hat{S}_{-} \), managing state decrementation. This ensemble represents a quantum approach to simulating financial market dynamics. }
\label{fig:FIG4}
\end{figure}

\subsection{Solution Architecture}\label{sec3}
The multi-SSQW framework leverages a dual-domain computational approach, involving a Parameterized Quantum Circuit (PQC) and a classical optimizer. The PQC is composed of n+1 qubits, one designated for coin space and the remainder for position space. This setup is employed to represent and emulate the distribution of short-term financial prices. To fine-tune these parameters in alignment with empirical data, the framework employs a classical optimizer. The optimizer implements the Constrained Optimization By Linear Approximation (COBYLA) algorithm to refine the trained results toward the target distribution. The COBYLA optimizer uses the mean-square error (MSE) and KL divergence as loss functions for an enhanced approach to the targeted distribution. This methodology allows for financial simulation and  preparing the probability data into a quantum state, effectively bridging the gap between classical finance models and quantum simulation.

The coin space of a multi-SSQW that performs a controlled motion of a walker on the position space  is similar to the ancilla qubit taking a controlled rotation in Eq.(\ref{eq:controlratation}). The goal is to optimize the coin parameters of a multi-SSQW to achieve the targeted distribution of the position space, and then we only compute the position space. We will accomplish this by using parameterized quantum circuits (PQC). The steps for this process are as follows:

\begin{itemize}
\item  Begin with a classical targeted data set $p=\{ p_{0},\ldots, p_{2^{N}-1} \} \in R$ sampled from a distribution of short-term financial prices.
\item  Multi-SSQW implementation uses an auxiliary qubit representing the coin space and N qubits representing the $2^{N}$ distributions in the position space.
\item Imply $\hat{W}_{1}, \hat{W}_{2}...\hat{W}_{k}$ operators or walkers on the quantum circuit and repeat $t$ steps.
\item Measure the state amplitudes of the position space and compute the trained distribution.
\item Update the coin parameters using the classical optimizer with the mean square error(MSE) and KL-divergence to quantify the difference between the trained probability distribution from the  targeted probability distribution.
\item Iterate  $n$ times until converge to the targeted distribution
\end{itemize}

\section{Results}

\subsection{Performances with daily return distributions of stocks}

We conducted an extensive simulation to gauge the effectiveness of the multi-SSQW framework with daily return distributions of various stocks. Our results indicate that this approach successfully leverages the advantages of quantum computation within the financial arena. 

A daily return distribution offers a statistical portrayal of a financial asset's daily returns, such as shares or commodities. It's the day-to-day value change in percentage terms for the asset. This distribution illustrates the frequency of different return values, enabling investors to evaluate the associated risks and potential returns of an investment. Typically presented as a histogram, the x-axis represents the return percentage, while the y-axis indicates probability frequency. To create a daily return distribution, we initially collect real stock data from Yahoo Finance every day over a specified duration(2022-01~2022-03). We then categorize these daily returns into 16 distinct groups based on their respective percentages. These data are then transformed into a frequency distribution bar chart to provide a more visual and intuitive understanding of the returns distribution. Afterward, we employ the multi-SSQW approach to simulate these outcomes, yielding a more realistic probability distribution of the market. Our study involves simulating the daily return distribution over a quarter of various stocks or indices. Initially, we performed simulations 100 times, experimenting with \textbf{num}(the number of walkers) within a range from 1 to 10 and $\textbf{step}=1$. This allows us to analyze the error statistically, observe the convergence behavior, and document the computation time for each scenario. Subsequently, we introduce random parameters to execute optimization simulations. These simulations enable us to generate the resultant data and the status of error convergence, thereby validating the efficacy and reliability of our approach. This structured and systematic procedure helps to provide a comprehensive understanding of the daily return distribution for different financial assets.

\begin{figure}[htb]
\centering
\subcaptionbox{}{\includegraphics[width=6.1cm]{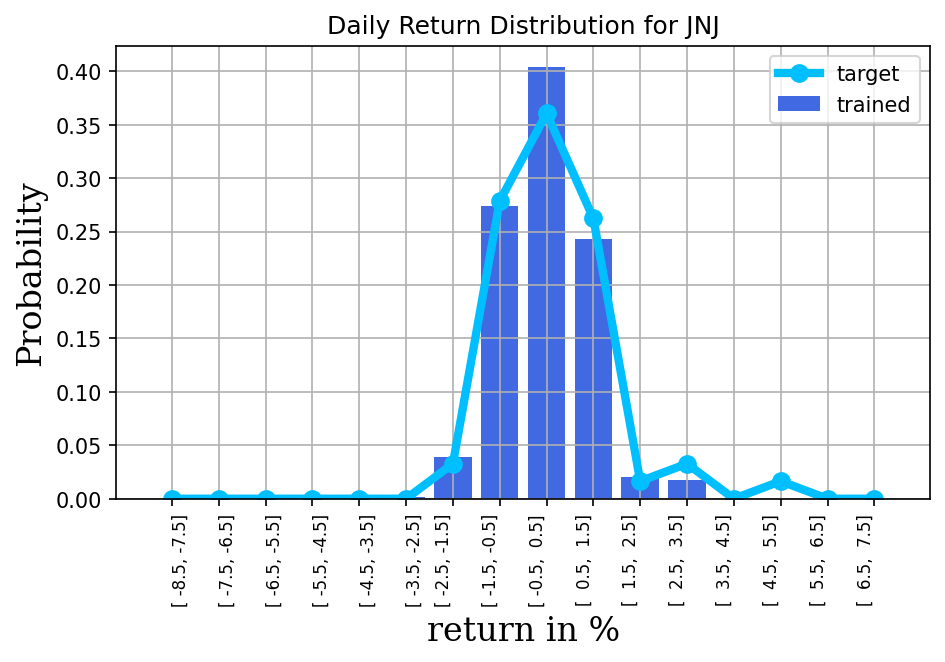}}
\hfill
\subcaptionbox{}{\includegraphics[width=6.1cm]{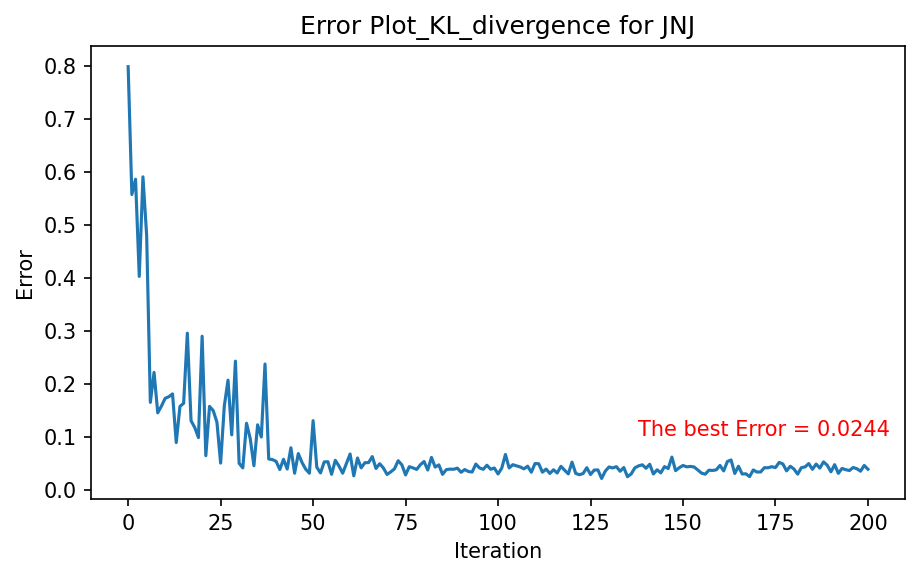}}
\hfill
\subcaptionbox{}{\includegraphics[width=6.1cm]{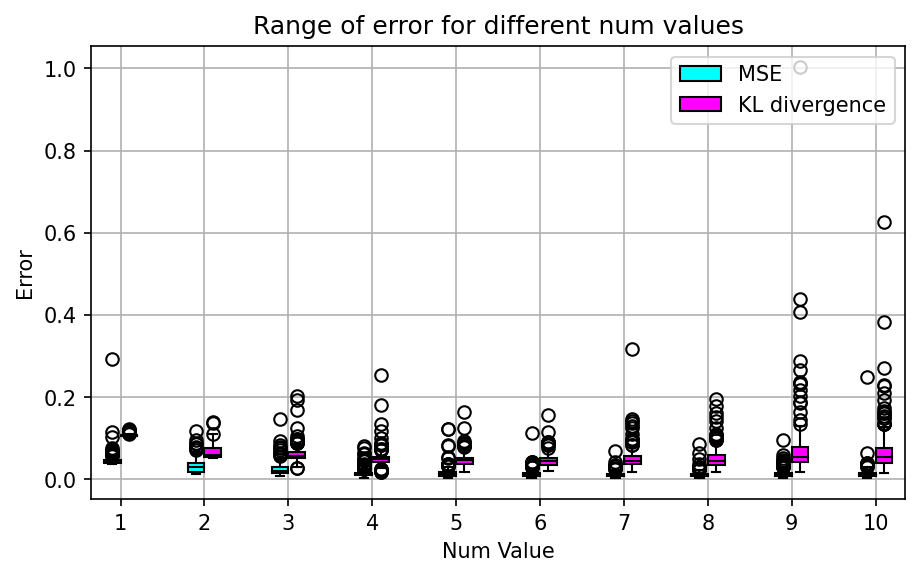}}
\hfill
\subcaptionbox{}{\includegraphics[width=6.1cm]{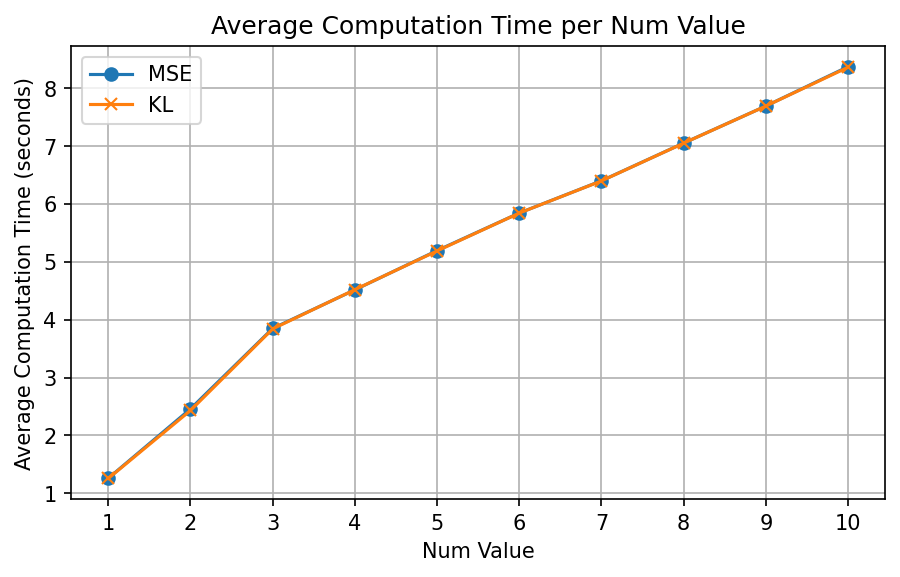}}
\caption{ The performances of the multi-SSQW method on Johnson and Johnson (JNJ) across various aspects: (a) the distribution of daily returns( Num = 4), (b) the progression of error, (c) a boxplot representation of the error, and (d) the average computational time required for the analysis.}
\label{fig:FIG5}
\end{figure}

In Fig. \ref{fig:FIG5}, we conducted an evaluation of the efficacy of the multi-SSQW method with respect to Johnson \& Johnson (JNJ). Our investigation involved the utilization of multi-SSQW to generate a simulation of the distribution of daily returns for JNJ over a quarter. The results indicate a strong correspondence when the number of walkers is set to 4, with the simulated and actual distributions closely aligning.

From a financial perspective, JNJ is a multinational corporation based in the United States that manufactures pharmaceuticals and consumer packaged goods and is often viewed as a defensive sector because the demand for healthcare products and services remains relatively steady, regardless of economic cycles. This characteristic makes JNJ an attractive option for conservative investors seeking stable returns. In our simulation, the findings display a strong alignment when the number of walkers is set to 4. This pattern suggests the prevalence of less sentiment-driven investors, which can be interpreted from a financial perspective as a tendency towards stability and predictability in the stock's performance.

\begin{figure}[htb]
\centering
\subcaptionbox{}{\includegraphics[width=6.1cm]{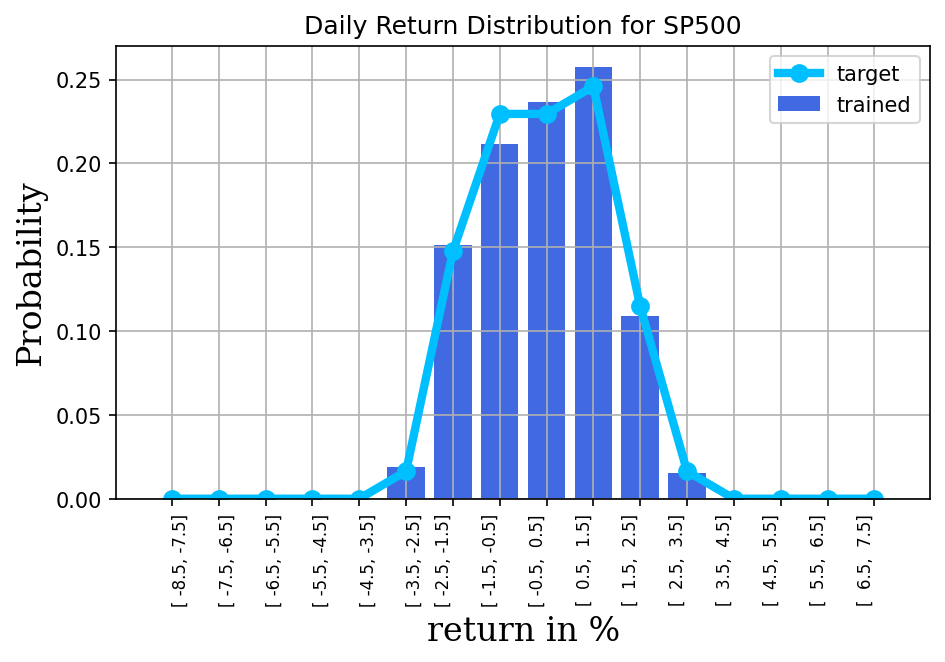}}
\hfill
\subcaptionbox{}{\includegraphics[width=6.1cm]{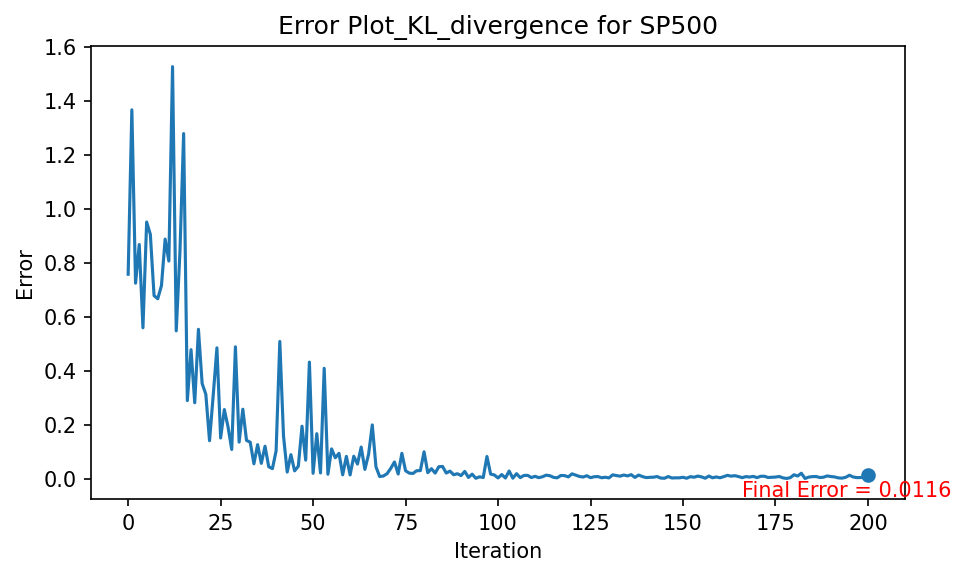}}
\hfill
\subcaptionbox{}{\includegraphics[width=6.1cm]{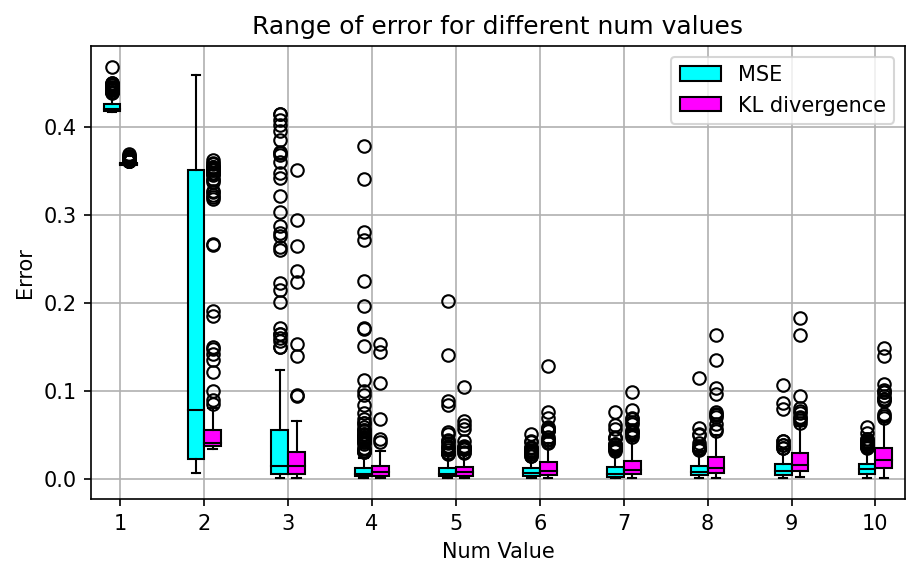}}
\hfill
\subcaptionbox{}{\includegraphics[width=6.1cm]{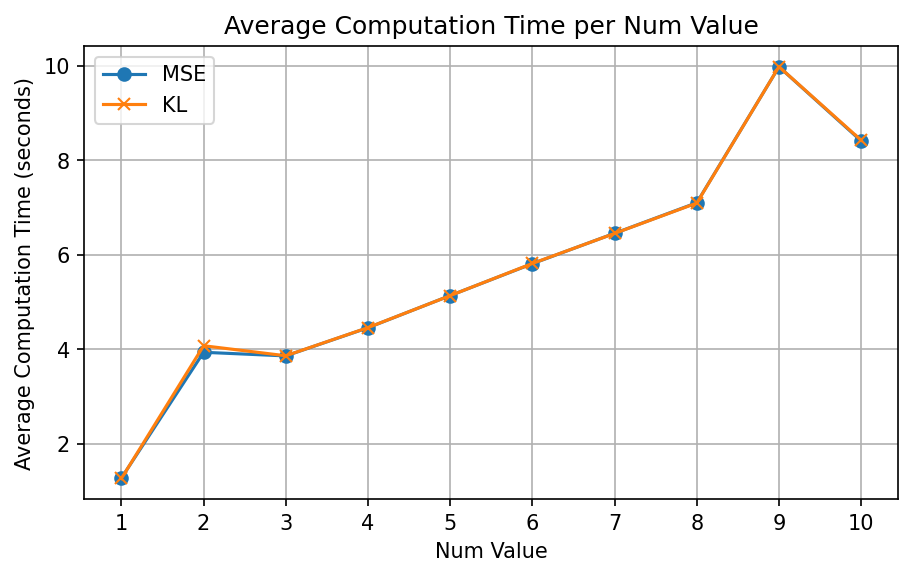}}
\caption{ The performances of the multi-SSQW method on SP500 across various aspects: (a) the distribution of daily returns(Num = 4), (b) the progression of error, (c) a boxplot representation of the error, and (d) the average computational time required for the analysis.}
\label{fig:FIG6}
\end{figure}

In Fig \ref{fig:FIG6}, we simulate the distribution of S\&P500. The Standard \& Poor's 500(S\&P500) is a stock market index that measures the stock performance of 500 large companies listed on stock exchanges in the United States. The index includes companies from all sectors of the economy. For large companies such as those in the S\&P 500, stable operations, good corporate governance, clear future planning, and strong performance all attract long-term investors. Given the relatively homogeneous investor base, fewer numbers are required for the simulation results to converge to an optimal outcome.

\begin{figure}[htb]
\centering
\subcaptionbox{}{\includegraphics[width=6.1cm]{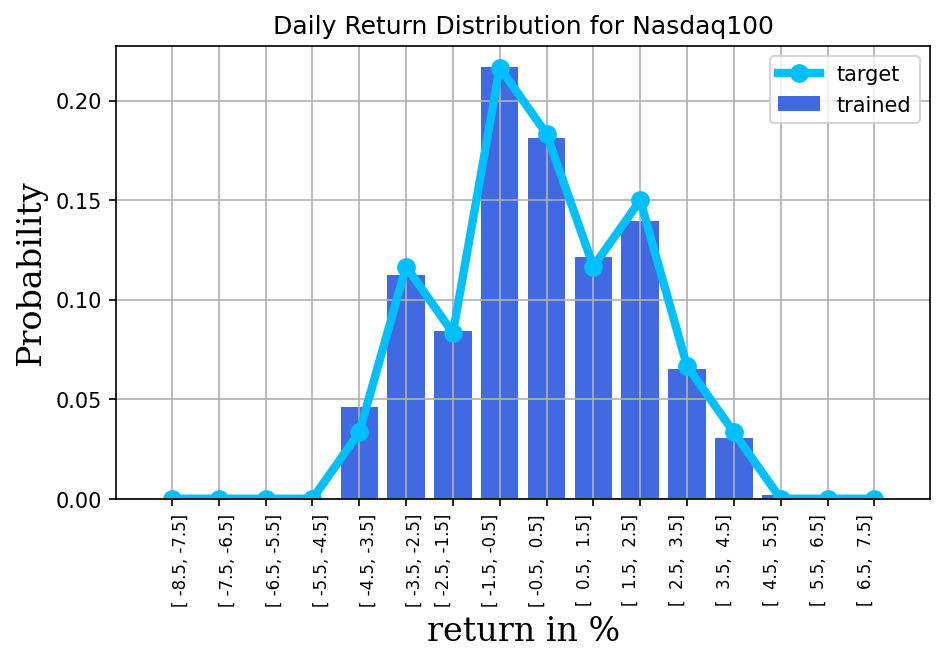}}
\hfill
\subcaptionbox{}{\includegraphics[width=6.1cm]{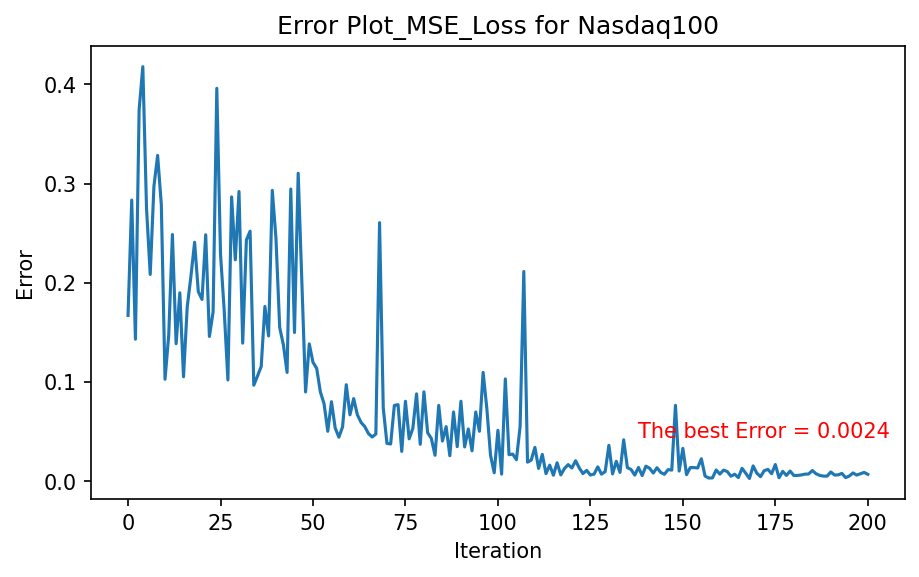}}
\hfill
\subcaptionbox{}{\includegraphics[width=6.1cm]{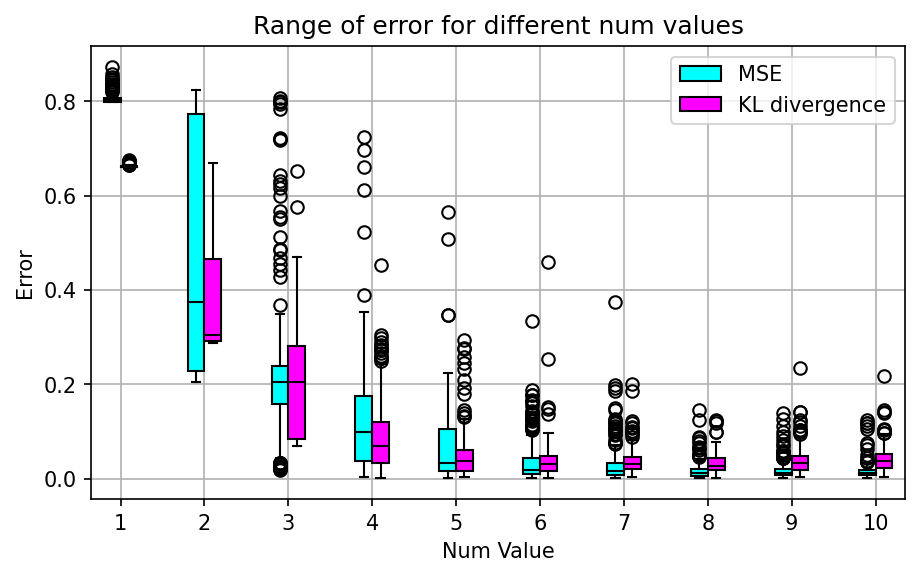}}
\hfill
\subcaptionbox{}{\includegraphics[width=6.1cm]{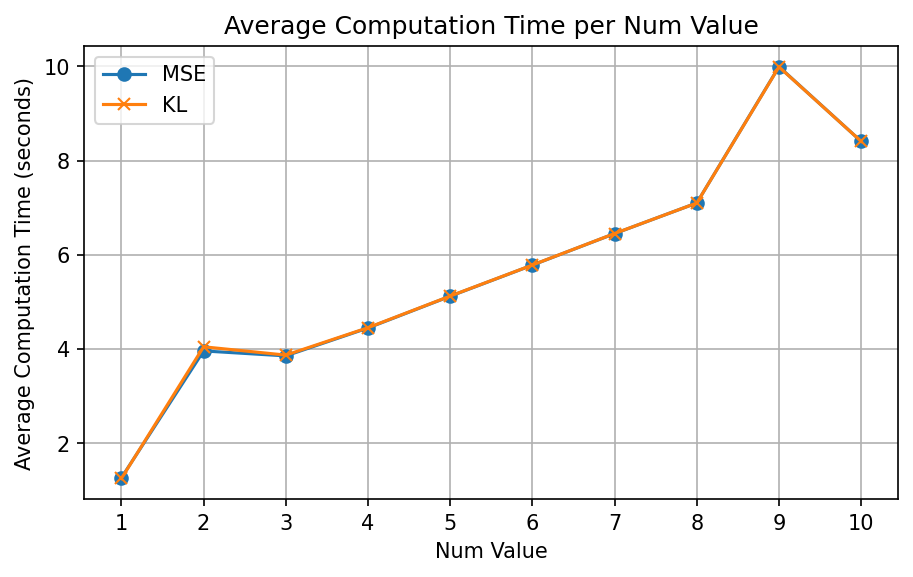}}
\caption{ The performances of the multi-SSQW method on Nasdaq100 across various aspects: (a) the distribution of daily returns(Num = 7), (b) the progression of error, (c) a boxplot representation of the error, and (d) the average computational time required for the analysis.}
\label{fig:FIG7}
\end{figure}

The development of the technology sector is highly dependent on innovation, which often makes the prospects of tech companies full of uncertainty. A breakthrough technology or product can swiftly disrupt the market dynamics and are more prone to sentiment. This tends to draw in a significant number of noise investors who do not trade on the basis of information and make irrational investment decisions. Consequently, this causes the stock prices of some companies to experience dramatic swings. Tech stocks are usually significantly influenced by market sentiment. In Fig.\ref{fig:FIG7}, the simulation shows the behavior could indicate a larger proportion of sentiment-driven investors. 

\begin{figure}[htb]
\centering
\subcaptionbox{}{\includegraphics[width=6.1cm]{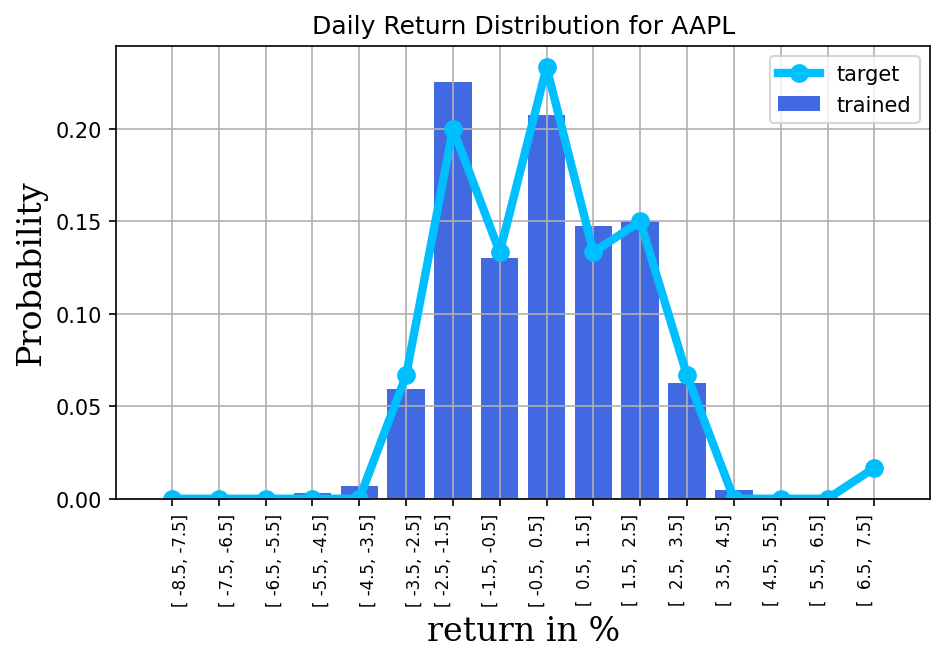}}
\hfill
\subcaptionbox{}{\includegraphics[width=6.1cm]{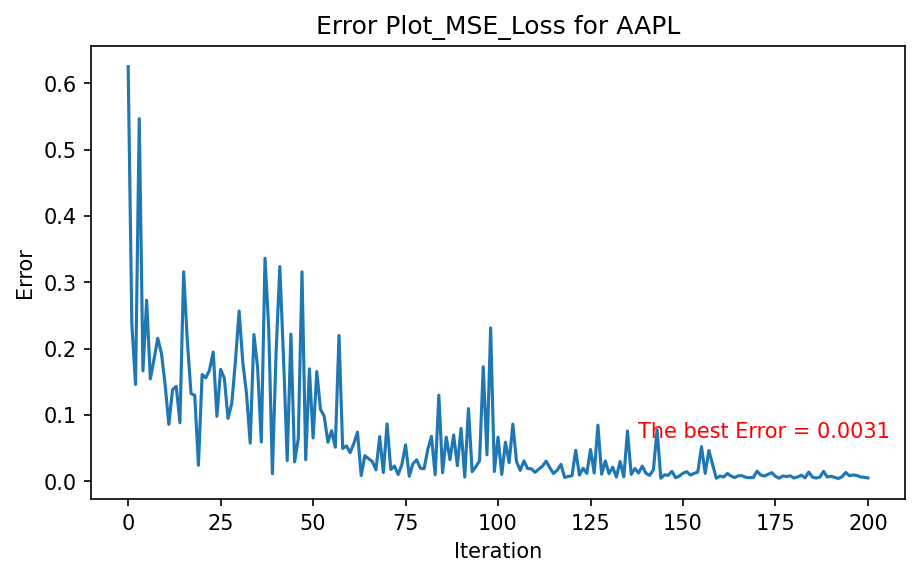}}
\hfill
\subcaptionbox{}{\includegraphics[width=6.1cm]{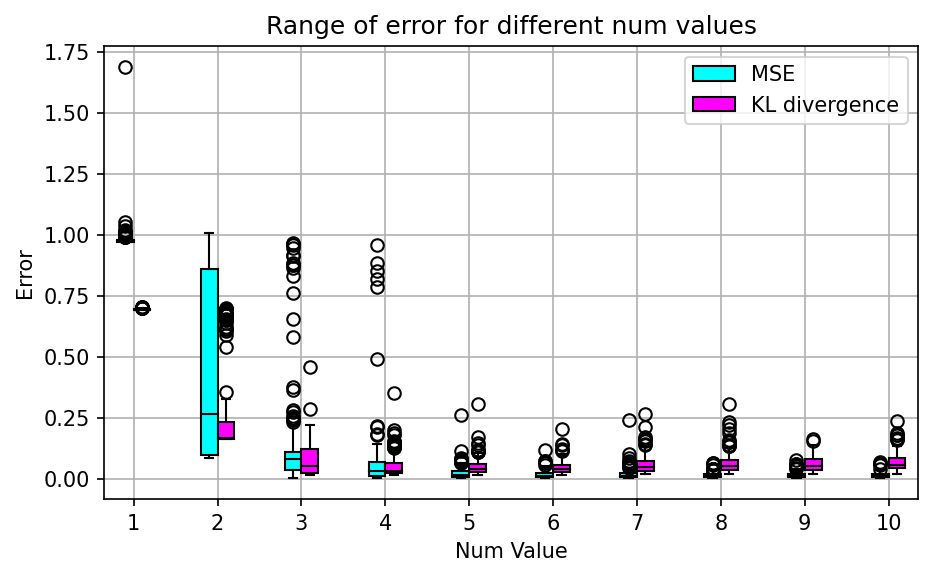}}
\hfill
\subcaptionbox{}{\includegraphics[width=6.1cm]{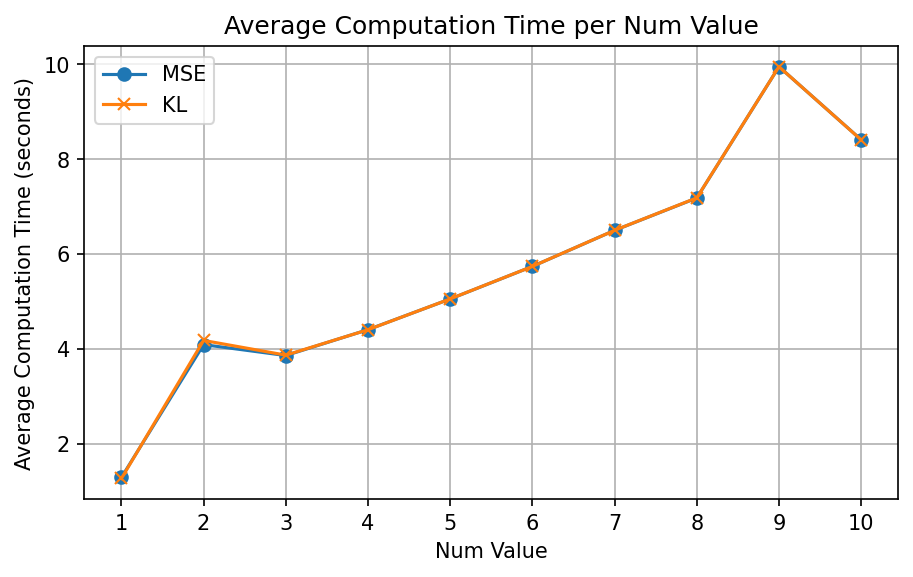}}
\caption{ The performances of the multi-SSQW method on APPLE (APPL) across various aspects: (a) the distribution of daily returns(Num = 7), (b) the progression of error, (c) a boxplot representation of the error, and (d) the average computational time required for the analysis.}
\label{fig:FIG8}
\end{figure}

\begin{figure}[htb]
\centering
\subcaptionbox{}{\includegraphics[width=6.1cm]{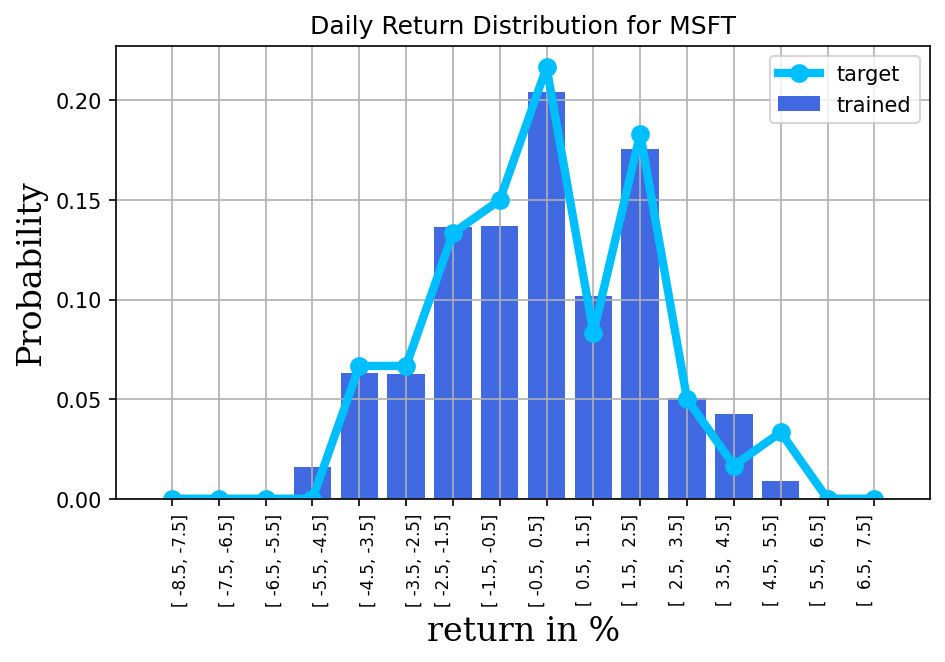}}
\hfill
\subcaptionbox{}{\includegraphics[width=6.1cm]{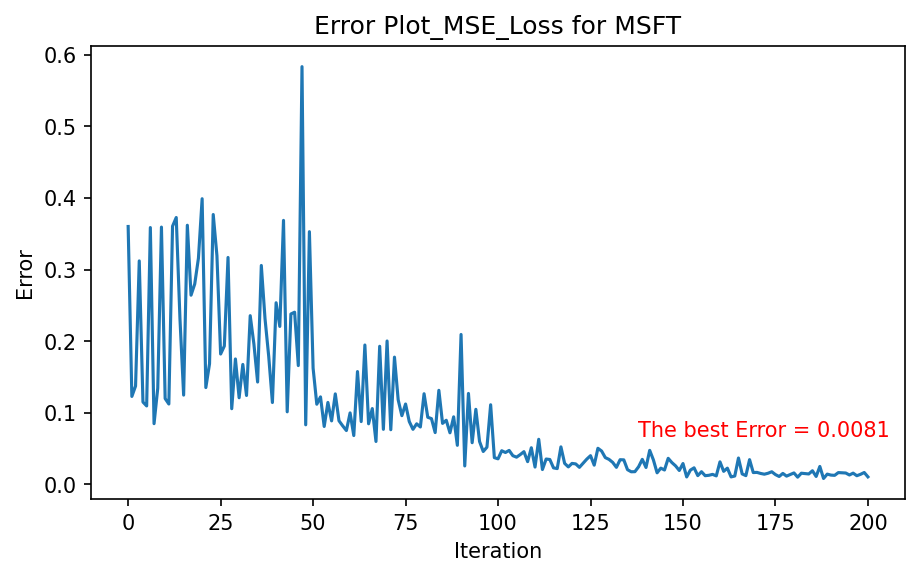}}
\hfill
\subcaptionbox{}{\includegraphics[width=6.1cm]{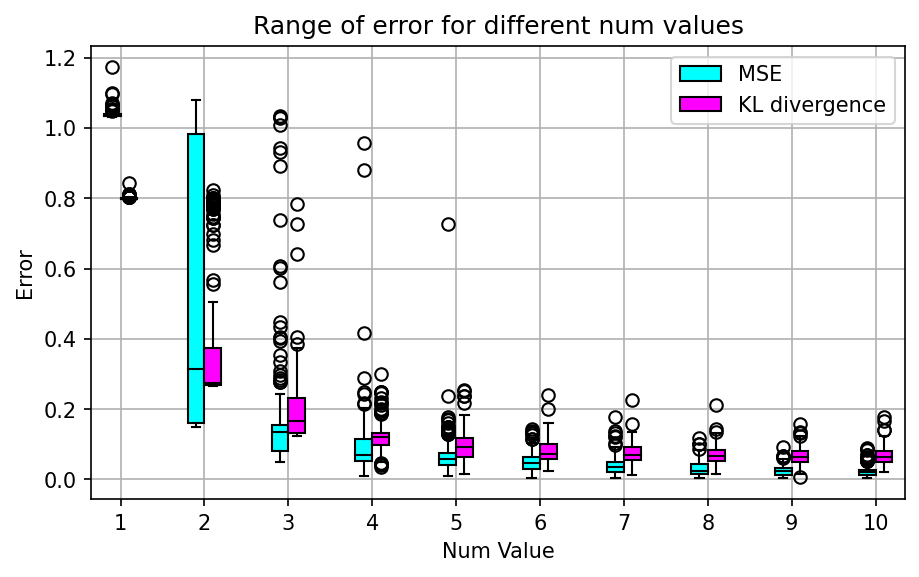}}
\hfill
\subcaptionbox{}{\includegraphics[width=6.1cm]{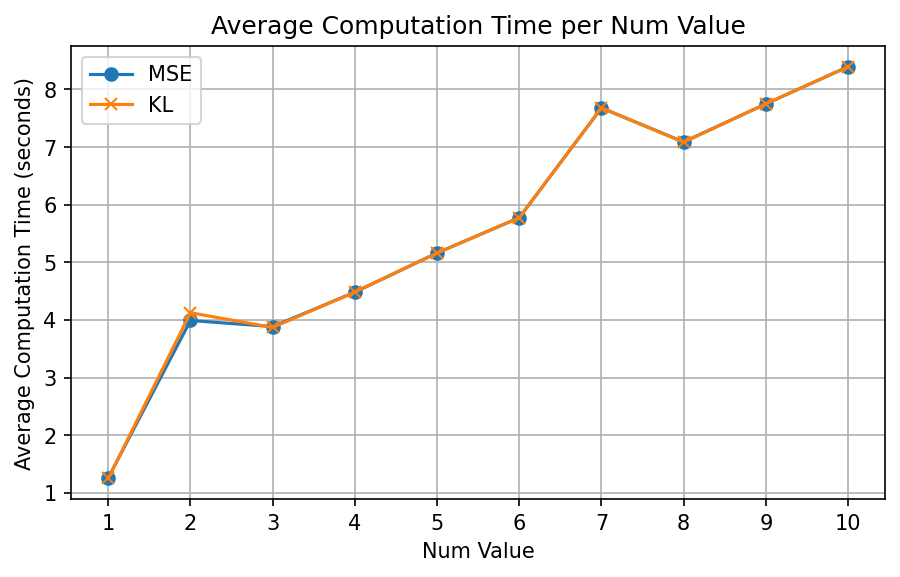}}
\caption{ The performances of the multi-SSQW method on Microsoft (MSFT) across various aspects: (a) the distribution of daily returns(Num = 7), (b) the progression of error, (c) a boxplot representation of the error, and (d) the average computational time required for the analysis.}
\label{fig:FIG9}
\end{figure}
Apple and Microsoft stand among the world's leading technology giants. They have a major influence on global technology trends and garners significant attention from investors globally. However, it is important to note that its distribution diverges from the typical normal distribution, escalating the complexity involved in its simulation. In Fig. \ref{fig:FIG8} and \ref{fig:FIG9}, we demonstrate the robust simulation capabilities of the multi-SSQW methodology, specifically applied to complex entities such as Apple and Microsoft.

In this section, we demonstrate the ability of multi-SSQW to function as an effective financial simulator. It is capable of accurately modeling intricate financial systems and providing reliable simulations. One of the highlights of this approach is its inherent capability to exhibit convergence and provide rapid results, which makes it a powerful tool in financial analytics and modeling. The boxplots reflect simulation outcomes that measure the multi-SSQW's fidelity in financial market modeling. They suggest that as the diversity of quantum walkers —representative of market participants— increases, the precision of the simulations improves, marked by lower MSE and KL divergence values. This relationship underscores the robustness of the multi-SSQW approach in capturing the complex dynamics of financial markets, providing a compelling tool for analysts. The methodology's reliable convergence indicates its utility in producing accurate market simulations, affirming its value in enhancing financial analysis and forecasting. In machine learning, integrating more parameters can boost accuracy without imposing a substantial computational load. Crucially, the method demonstrates steady convergence, highlighting this approach's efficacy.

\subsection{Performances of the binomial distribution}

The binomial distribution is a probability distribution that describes the number of successes in a fixed number of independent Bernoulli trials with the same probability of success. A common example of a binomial distribution is a coin toss, where the outcome can be either heads (success) or tails (failure), and each toss of the coin is an independent event. 

\begin{figure}[htb]
\centering
\subcaptionbox{}{\includegraphics[width=6.1cm]{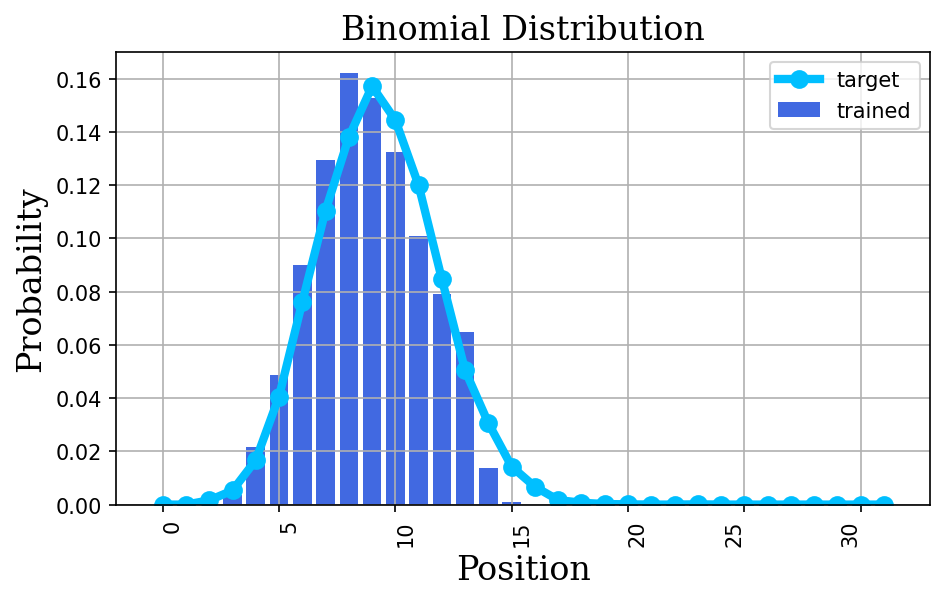}}
\hfill
\subcaptionbox{}{\includegraphics[width=6.1cm]{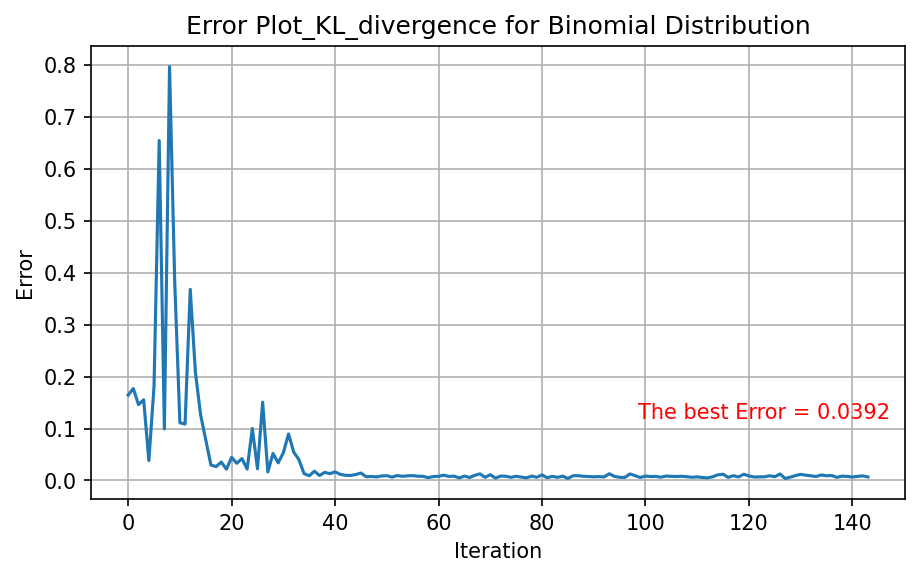}}
\hfill
\subcaptionbox{}{\includegraphics[width=6.1cm]{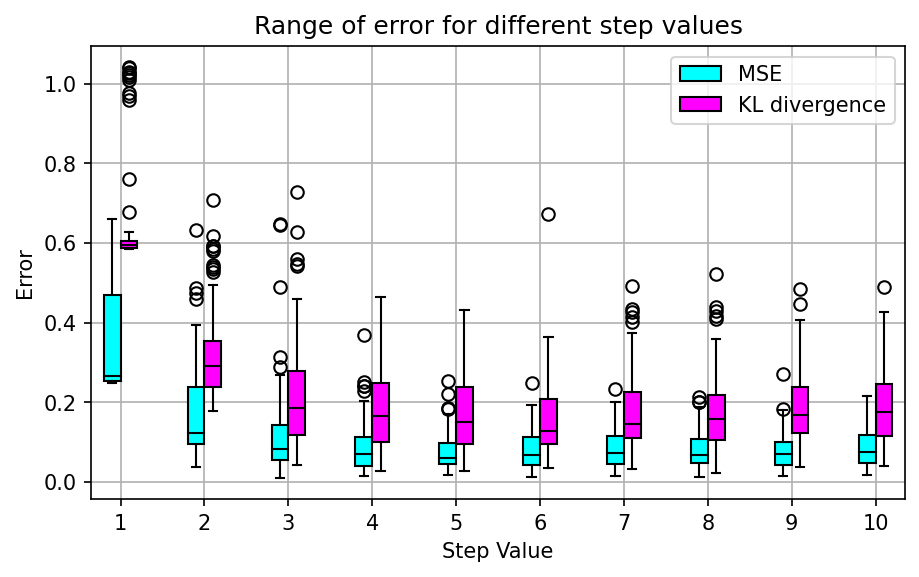}}
\hfill
\subcaptionbox{}{\includegraphics[width=6.1cm]{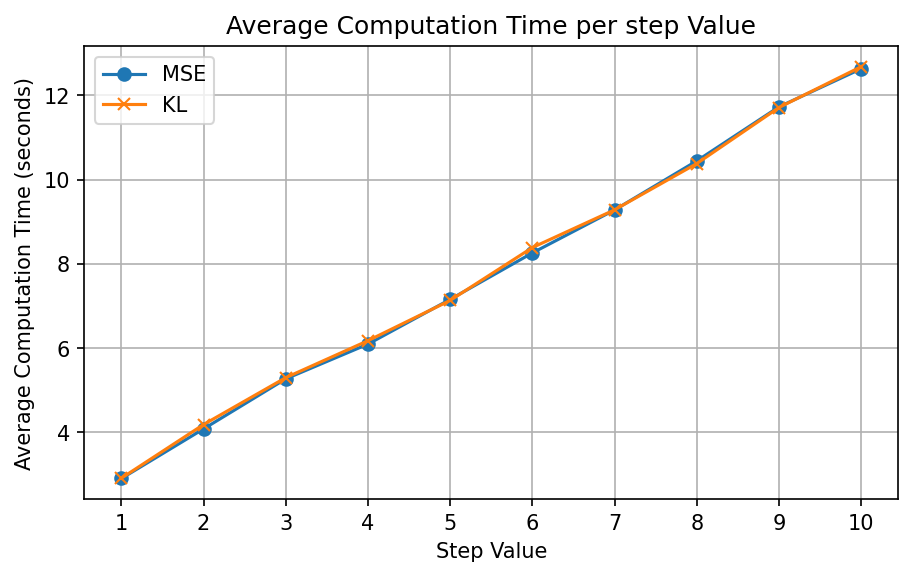}}
\caption{ The performances of the binomial distribution across various aspects: (a) the binomial distribution ( p = 0.3, n= 31, num =2, step = 3), (b) the progression of error, (c) a boxplot representation of the error, and (d) the average computational time required for the analysis.}
\label{fig:FIG10}
\end{figure}

\begin{figure}[htb]
\centering
\subcaptionbox{}{\includegraphics[width=6.1cm]{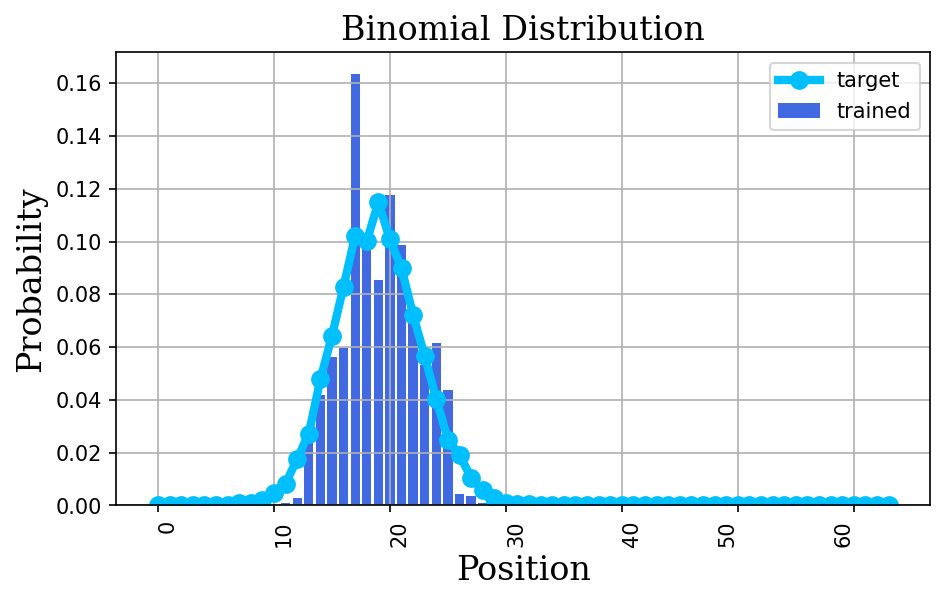}}
\hfill
\subcaptionbox{}{\includegraphics[width=6.1cm]{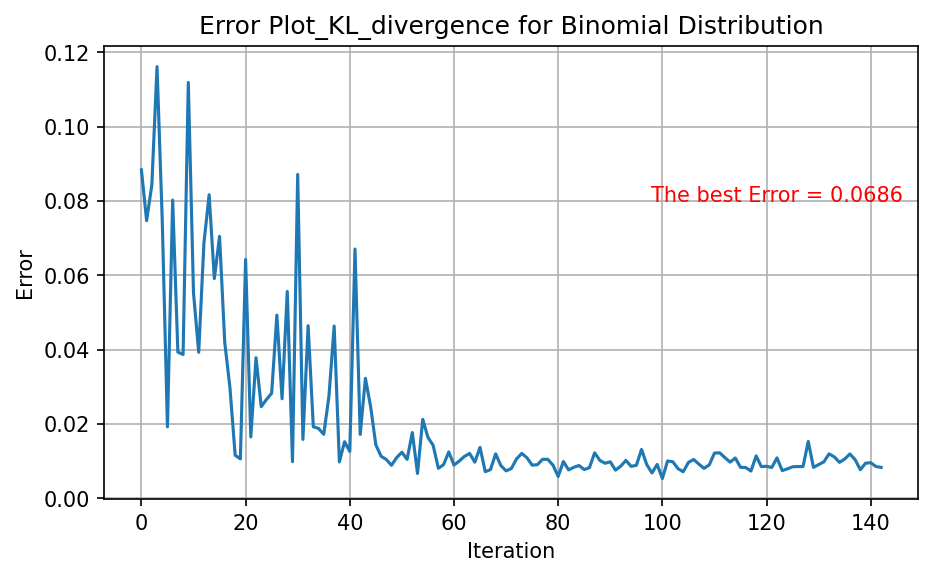}}
\hfill
\subcaptionbox{}{\includegraphics[width=6.1cm]{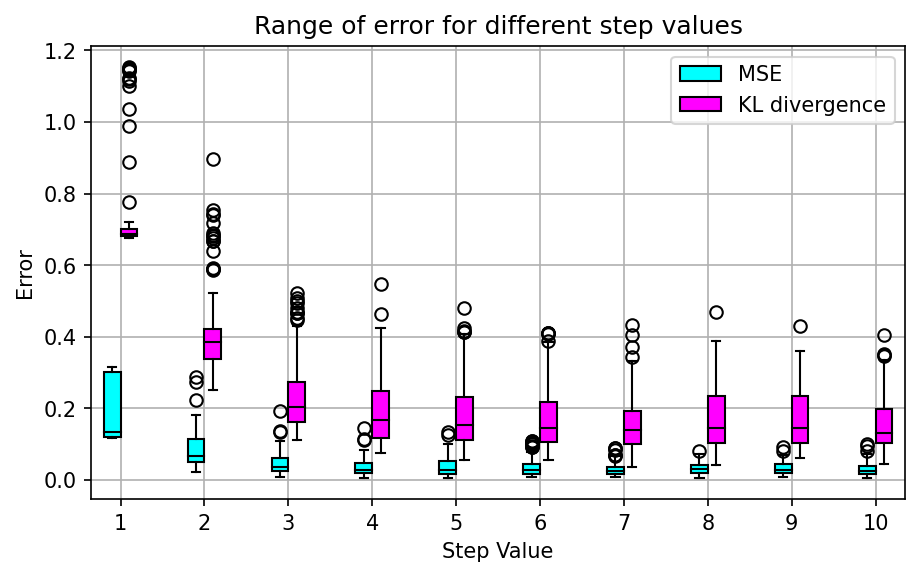}}
\hfill
\subcaptionbox{}{\includegraphics[width=6.1cm]{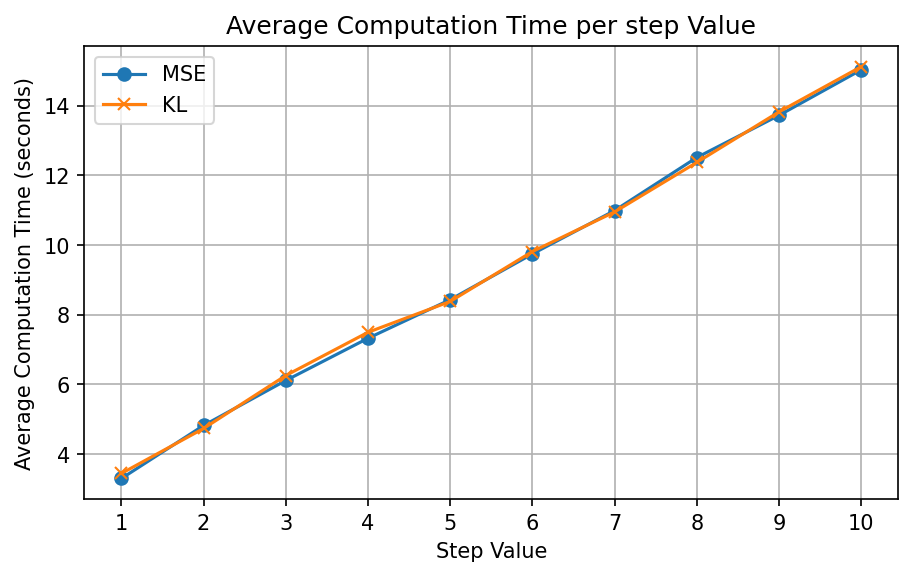}}
\caption{ The performances of the binomial distribution across various aspects: (a) the binomial distribution ( p = 0.3, n= 63, num = 3, step = 5), (b) the progression of error, (c) a boxplot representation of the error, and (d) the average computational time required for the analysis.}
\label{fig:FIG11}
\end{figure}

\begin{figure}[htb]
\centering
\subcaptionbox{}{\includegraphics[width=6.1cm]{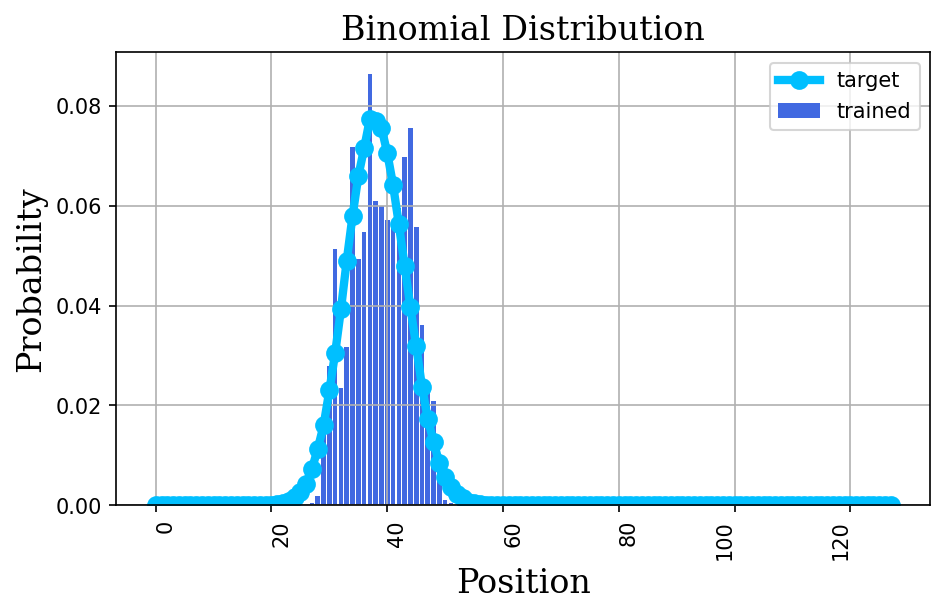}}
\hfill
\subcaptionbox{}{\includegraphics[width=6.1cm]{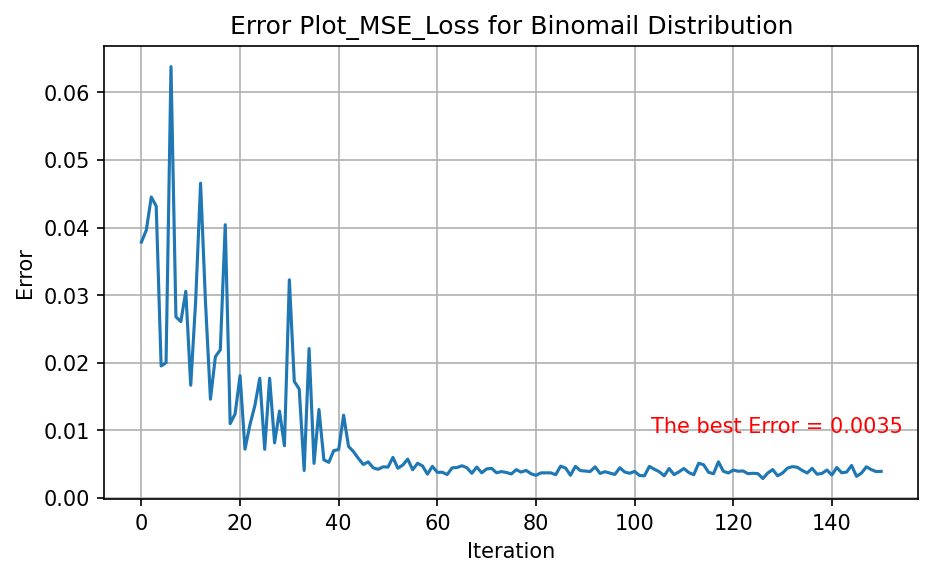}}
\hfill
\subcaptionbox{}{\includegraphics[width=6.1cm]{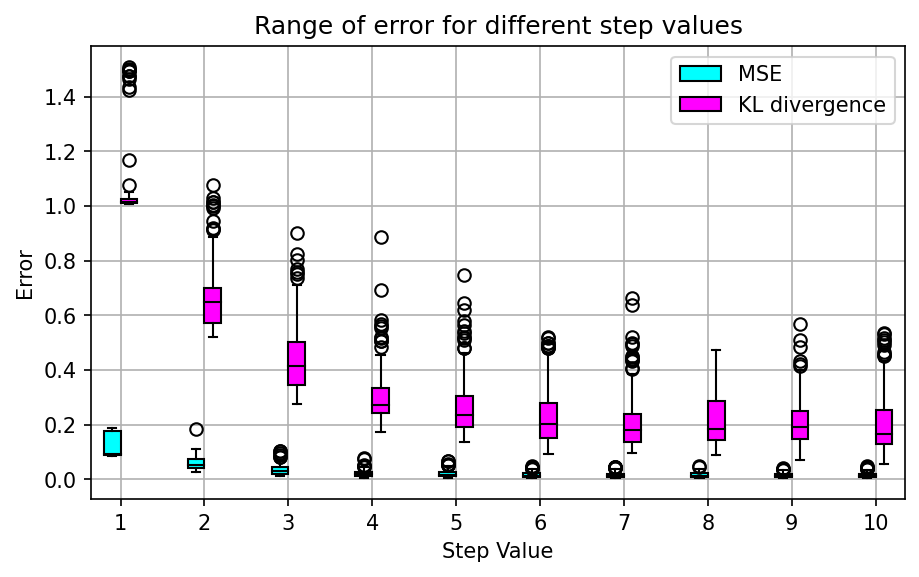}}
\hfill
\subcaptionbox{}{\includegraphics[width=6.1cm]{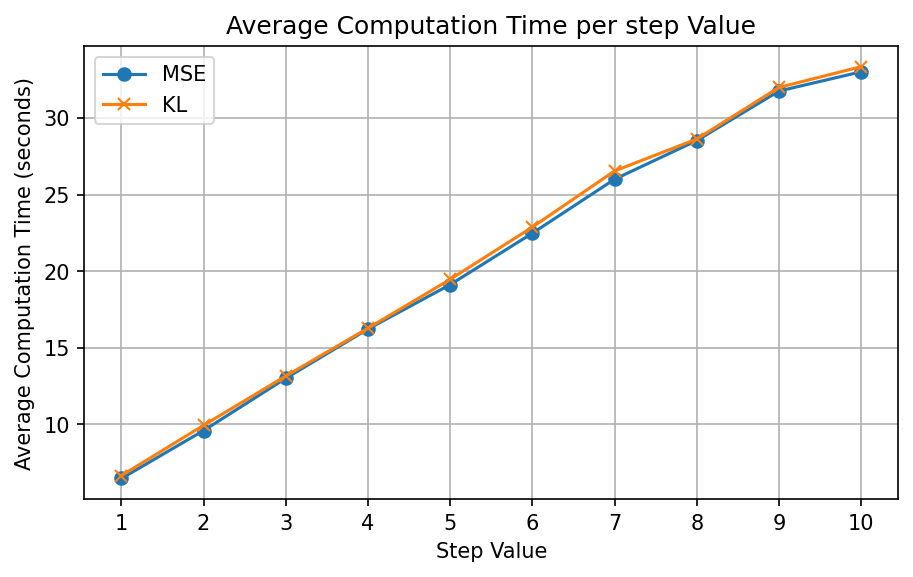}}
\caption{ The performances of the binomial distribution across various aspects: (a) the binomial distribution ( p = 0.3, n= 127, num = 3, step = 10), (b) the progression of error, (c) a boxplot representation of the error, and (d) the average computational time required for the analysis.}
\label{fig:FIG12}
\end{figure}

Probability theory\cite{stroock_2010,kallenberg2002foundations}, the mathematical study of randomness, is built upon the concept of probability distributions and the random variables they describe. Every distribution has a specific application and is characterized by certain parameters which help to define the shape and probabilities of the distribution. We demonstrate the utilization of multi-SSQW for the simulation of binomial distributions, shown in Fig. \ref{fig:FIG10}-\ref{fig:FIG12}. These are accomplished by adjusting the coin operator coefficient and the walker count to mirror the success probability and fine-tuning the number of control steps to match the characteristics of the binomial distribution. We present three distinct binomial distributions, each with a success probability of 0.3, undergoing 31, 63, and 127 trials, respectively. The outcomes, derived via the multi-SSQW approach, validate our ability to use two quantum walks, incorporating six parameters each and a suitable number of steps, to depict the binomial distribution. The multi-SSQW approach enables us to procure a statistical approximation of the binomial distribution with control steps of 3, 5, and 10, respectively.

\subsection{Application: European call option price }

The logarithmic normal distribution is a probability distribution of a random variable whose logarithm is normally distributed. One of the main properties of the log-normal distribution is that the values are skewed to the right, meaning that the tail of the distribution is on the right side, representing large values. It also has thicker tails than a normal distribution, meaning extreme events are more likely. This is why it helps model variables that fluctuate widely, such as stock prices. The log-normal distribution also can be used to price options by assuming that the underlying asset's price follows a geometric Brownian motion, which means that the logarithm of the asset's price follows a normal distribution. This assumption is the basis of the Black-Scholes model, which is a widely used method for pricing options.  This is evidenced by its practical application in financial derivative pricing and the simulation of financial markets. In the following, we demonstrated that the multi-SSQW scheme enables the exploitation of the potential quantum advantage in finance, such as European call option pricing.

The Black-Scholes (BS) model\cite{BSmodel}  is widely used in the financial industry to value options and other financial derivatives. It is used to determine the theoretical value of an option using specific parameters such as the underlying asset's price$(S)$, strike price$(K)$, the time to expiration$(T)$, the risk-free interest rate$(r)$ and the volatility$(vol)$. The BS model calculates the option value by assuming that the underlying asset's price follows a geometric Brownian motion, a continuous-time stochastic process. Considering this, the BS model can calculate the probability distribution of the underlying asset's price at expiration.

\begin{figure}[htb]
\centering
\subcaptionbox{}{\includegraphics[width=6.1cm]{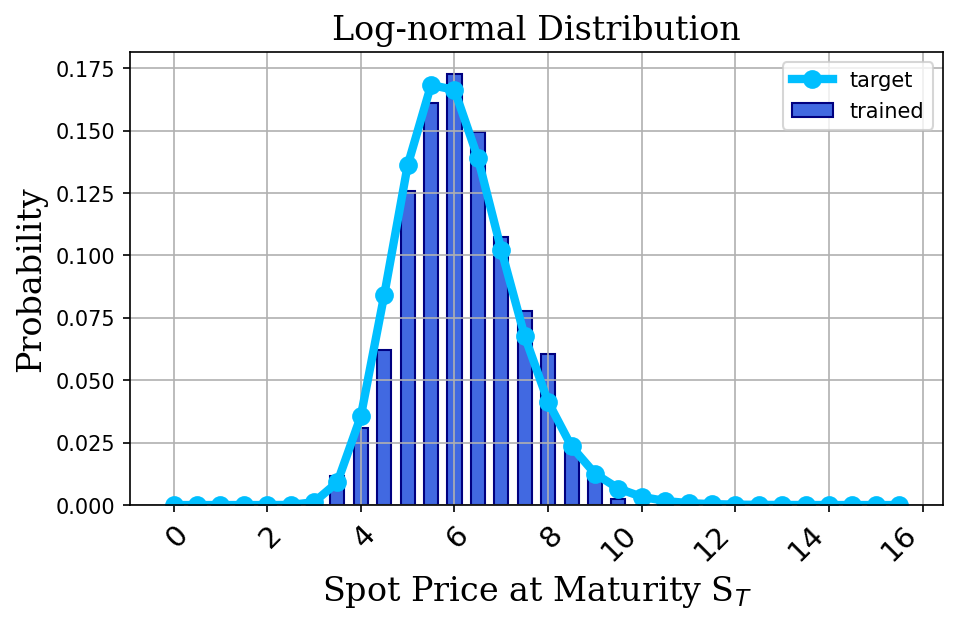}}
\hfill
\subcaptionbox{}{\includegraphics[width=6.1cm]{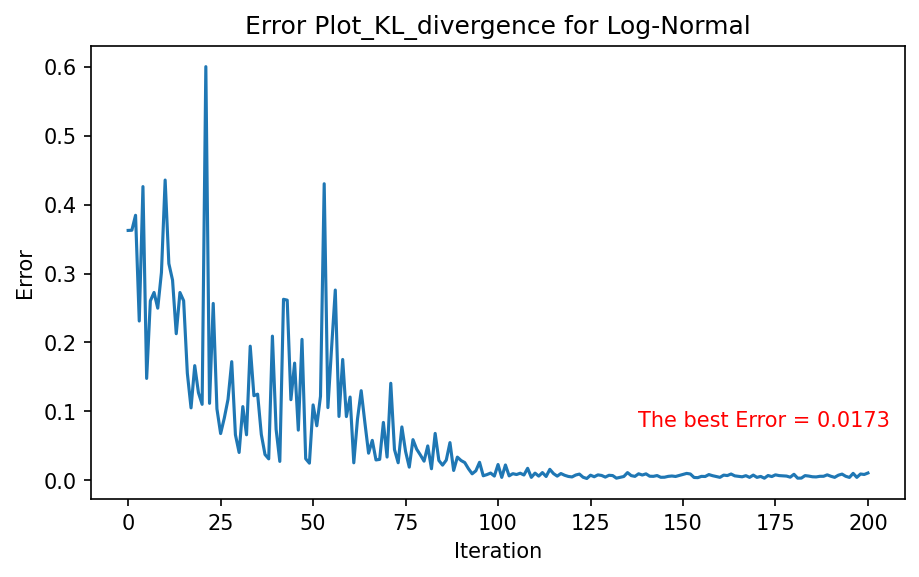}}
\caption{ The performances of the log-normal distribution (a) the log-normal distribution ( $Num = 3, step = 4$), (b) the progression of error }
\label{fig:FIG13}
\end{figure}

We approximate the trained probability distribution, leveraging multi-SSQW with $num=3$ and $step=4$, and compare it with the targeted probability distribution for the parameters  by plotting them together, as shown in Fig. \ref{fig:FIG13}. Now, the trained probability distribution can be used to evaluate the expectation value of the option's payoff function. We can see that the analytically calculated expected payoff is 0.1739 when using the targeted distribution, and 0.1460 when using the trained distribution. Therefore, when the trained probability distribution is more accurate, resulting in a more favorable outcome after calculation.

Drawing on previous experience, we concurrently adjust the $num$ and $step$ parameters to model the log-normal distribution. We construct an approximation of the trained probability distribution, employing multi-SSQW with $num =3$ and $step =4$. This approximation is juxtaposed against the targeted probability distribution for parameters $S_{0}=6$,$K=7$, $vol = 0.4$, $r = 0.04$ and $T=90$. The comparison is visualized in Fig. \ref{fig:FIG13}. The trained probability distribution is subsequently used to evaluate the expected value of the option's payoff function. The analytical computation reveals an expected payoff of 0.1739 when using the targeted distribution and 0.1460 when employing the trained distribution. Hence, the more precise the trained probability distribution, the more favorable the result after calculation.

\section{Discussions}\label{sec4}

In our research, we present preliminary theoretical exploration and practical applications that illustrate the effectiveness of our approach in conducting quantum financial simulations and quantum state preparation. Our strategy harnesses the unique strengths of quantum computation, and the multi-SSQW can accurately model intricate financial distributions and scenarios. This offers fresh insights and tools for financial analysis, decision-making, and quantum state preparation. The multi-SSQW algorithm offers several notable benefits:

\textbf{Flexibility in Modeling}: Multi-SSQW boasts exceptional adaptability to modeling complex systems, including those in the financial domain. It exhibits remarkable versatility by enabling adjustments to variables such as the number of walkers and steps and optimizing parameters in the coin space. This adaptability allows it to accurately depict a wide range of complex systems, making it a potent and flexible tool for applications in financial simulations and beyond.

\textbf{Stable Convergence}: A vital characteristic of the multi-SSQW is its stable convergence - a critical requirement for a resilient financial simulator. This feature ensures that the quantum walks consistently attain a stable state over time. Our results demonstrate that multi-SSQW can produce more accurate simulations. For instance, when approximating the binomial and log-normal distributions, we found that the error could be effectively minimized by increasing the number of walkers.

\textbf{Efficient Computation}: Multi-SSQW leverages efficiency by skillfully navigating the complex quantum state space. The simultaneous progression of multiple walkers in this space fosters swift exploration and convergence to the targeted state or solution. The fundamental principles of quantum computation, quantum superposition, and entanglement enhance this capability, permitting parallel processing and exponential computational speed-ups compared to classical approaches. This means the multi-SSQW not only operates faster but also utilizes resources in a markedly more efficient manner, offering a significant advantage in tackling complex problems quickly and effectively.

Further, by appropriately tuning the number of walkers, steps, and coin operator parameters, multi-SSQW can be optimized for swift convergence, thereby enhancing efficiency. It is important to note, however, that achieving efficiency necessitates a deep understanding of system dynamics and careful selection and tuning of parameters, a potentially challenging task.

There appears to be an inherent trade-off between the quantum circuit's complexity and the optimizer's efficiency in the Variational Quantum Algorithms (VQA). As we increase the depth and intricacy of our quantum
circuits, the state space we can explore expands, potentially leading to better solutions. However, this also means
that the landscape of the cost function can become more intricate, possibly making optimization more challenging.
Recognizing and managing this balance is crucial to harnessing the full potential of VQA and achieving practical and
meaningful results. Using a multi-SSQW quantum circuit to simulate financial stock distributions can be seen as employing well-orchestrated circuits. The multi-SSQW approach is designed to navigate complex quantum state spaces efficiently, aiming for rapid exploration and convergence to a desirable state or solution through the controlled evolution of multiple walkers. The advantages of multi-SSQW, such as its rapid modeling and prediction capacity, make it highly beneficial for financial simulations in fast-paced and volatile financial markets.

In summary, using quantum computing techniques, the multi-SSQW model is promising for advancing our grasp of market dynamics. Anticipated research will focus on enriching the model’s forecasting prowess for stock price trends by merging it with
macroeconomic data and assessing the nuanced interactions between different stocks using entanglement to develop joint probability distributions. These efforts aim to yield a more intricate understanding of the financial markets’ interconnectivity.

\section{ACKNOWLEDGMENTS}
We thank IBM Quantum Hub at NTU for providing computational
resources and accesses for conducting the real quantum device experiments. We acknowledges support from National Science and technology council, Taiwan  under
Grants NSTC 112-2119-M-033-001, by the research project Applications of quantum computing in optimization and finances.


\begin{thebibliography}{47}

\bibitem{Shor1994}
P.W. Shor, "Algorithms for quantum computation: discrete logarithms and factoring," Proceedings 35th Annual Symposium on Foundations of Computer Science, Santa Fe, NM, USA, 124-134 (1994).
\url{https://doi.org/10.1109/SFCS.1994.365700}
\bibitem{Feynman1982}
R.P. Feynman, "Simulating physics with computers," International Journal of Theoretical Physics, \textbf{21}, 467-488 (1982).
\url{https://doi.org/10.1007/BF02650179}
\bibitem{Trabesinger2012}
A. Trabesinger, "Quantum simulation," Nature Phys, \textbf{8}, 263 (2012).
\url{https://doi.org/10.1038/nphys2258}

\bibitem{Bachelier:RW}
Louis Bachelier, Mark Davis, Alison Etheridge, "Louis Bachelier's Theory of Speculation: The Origins of Modern Finance," Princeton: Princeton University Press, (2007).
\url{https://doi.org/10.1515/9781400829309}

\bibitem{Fama_EMH}
Eugene F.Fama, "Quantum amplitude amplification and estimation 'Efficient Capital Markets: A Review of Theory and Empirical Work'," The Journal of Finance, \textbf{25}(2), 383–417 (1970).
\url{https://doi.org/10.2307/2325486}

\bibitem{9222275}
Daniel J. Egger, Claudio Gambella, Jakub Marecek, et al., "Quantum Computing for Finance: State-of-the-Art and Future Prospects," IEEE Transactions on Quantum Engineering, \textbf{1}, 1-24 (2020).
\url{https://doi.org/10.1109/TQE.2020.3030314}

\bibitem{Herman2022}
Dylan Herman, Cody Googin, Xiaoyuan Liu, et al., "A Survey of Quantum Computing for Finance," arXiv, (2022).
\url{https://doi.org/10.48550/ARXIV.2201.02773}

\bibitem{Brassard_2002}
Gilles Brassard and Peter Høyer and Michele Mosca and Alain Tapp, "Quantum amplitude amplification and estimation," American Mathematical Society, \textbf{305}, 53-74 (2002).
\url{https://doi.org/10.1090/conm/305/05215}

\bibitem{Rebentrost2018}
Patrick Rebentrost, Brajesh Guptand Thomas R. Bromley, "Quantum computational finance: Monte Carlo pricing of financial derivatives," Phys. Rev. A, \textbf{98}, 022321 (2018).
\url{https://doi.org/10.1103/PhysRevA.98.022321}

\bibitem{Zoufal:qGAN2019}
C. Zoufal, A. Lucchi and S. Woerner, "Quantum Generative Adversarial Networks for learning and loading random distributions," npj Quantum Inf, \textbf{5}, 103 (2019).
\url{https://doi.org/10.1038/s41534-019-0223-2}


\bibitem{Stamatopoulos_2020}
Nikitas Stamatopoulos, Daniel J. Egger, Yue Sun, Christa Zoufal, Raban Iten, Ning Shen, and Stefan Woerner, "Option Pricing using Quantum Computers," Quantum, \textbf{4}, 291 (2020).
\url{https://doi.org/10.22331/q-2020-07-06-291}

\bibitem{Stamatopoulos_2022}
Nikitas Stamatopoulos, Guglielmo Mazzola, Stefan Woerner, and William J. Zeng, "Towards Quantum Advantage in Financial Market Risk using Quantum Gradient Algorithms," Quantum, \textbf{6}, 770 (2022).
\url{https://doi.org/10.22331/q-2022-07-20-770}

\bibitem{Dong_An_2021}
Dong An, Noah Linden, JinPing Liu, Ashley Montanaro, Changpeng Shao, and Jiasu Wang, "Quantum-accelerated multilevel Monte Carlo methods for stochastic differential equations in mathematical finance," Quantum, \textbf{5}, 481 (2021).
\url{https://doi.org/10.22331/q-2021-06-24-481}

\bibitem{Chakrabarti_2021}
Shouvanik Chakrabarti, Rajiv Krishnakumar, Guglielmo Mazzola, Nikitas Stamatopoulos, Stefan Woerner, and William J. Zeng, "A Threshold for Quantum Advantage in Derivative Pricing," Quantum, \textbf{5}, 463 (2021).
\url{https://doi.org/10.22331/q-2021-06-01-463}

\bibitem{HHL:2009}
A. W. Harrow, A. Hassidim, and S. Lloyd, "Quantum algorithm for linear systems of equations," Phys. Rev. Lett., \textbf{103}, 150502 (2009).
\url{https://doi.org/10.1103/PhysRevLett.103.150502}

\bibitem{PCA:2014}
S. Lloyd, M. Mohseni, and P. Rebentrost, "Quantum principal component analysis," Nature Phys, \textbf{10}, 631-633 (2014).
\url{https://doi.org/10.1038/nphys3029}

\bibitem{QSVM:2019}
V. Havlíček, A.D. Córcoles, K. Temme, et al., "Supervised learning with quantum-enhanced feature spaces," Nature, \textbf{567}, 209-212 (2019).
\url{https://doi.org/10.1038/s41586-019-0980-2}

\bibitem{QC2021}
Chien-Hung Cho, Chih-Yu CHen, Kuo-Chin Chein, et al., "Quantum computation: Algorithms and Applications," Chinese Journal of Physics, \textbf{72}, 248-269 (2021).
\url{https://doi.org/10.1016/j.cjph.2021.05.001}

\bibitem{Montanaro2016}
A. Montanaro, "Quantum simulation," npj Quantum Information, \textbf{2}, 15023 (2016).
\url{https://doi.org/10.1038/npjqi.2015.23}

\bibitem{PhysRevLett.122.020502}
Yuval R. Sanders, Guang Hao Low, Artur Scherer, and Dominic W. Berry, "Black-Box Quantum State Preparation without Arithmetic," Phys. Rev. Lett., \textbf{122}, 020502 (2019).
\url{https://doi.org/10.1103/PhysRevLett.122.020502}

\bibitem{Choi2023}
J. Choi, A.L. Shaw, I.S. Madjarov, et al., "Preparing random states and benchmarking with many-body quantum chaos," Nature, \textbf{613}, 468-473 (2023).
\url{https://doi.org/10.1038/s41586-022-05442-1}

\bibitem{Nielsen2000}
M. A. Nielsen, and I. L. Chuang, "Quantum Computation and Quantum Information," Cambridge University Press (2010).
\url{https://doi.org/10.1017/CBO9780511976667}

\bibitem{Guise2018}
Hubert de Guise, Olivia Di Matteo, and Luis L. Sánchez-Soto, "Simple factorization of unitary transformations," Phys. Rev. A, \textbf{97}, 022328 (2018).
\url{https://doi.org/10.1103/PhysRevA.97.022328}

\bibitem{Grover:loading2002}
L. Grover and T. Rudolph, "Creating superpositions that correspond to efficiently integrable probability distributions," arXiv:quant-ph/0208112v1 (2002).
\url{https://doi.org/10.48550/arXiv.quant-ph/0208112}

\bibitem{Rocchetto2018QSP}
A. Rocchetto, E. Grant, S. Strelchuk, et al., "Learning hard quantum distributions with variational autoencoders," npj Quantum Information, \textbf{4}, 28 (2018).
\url{https://doi.org/10.1038/s41534-018-0077-z}

\bibitem{Kalayn:Loading2022}
Kalyan Dasgupta and Binoy Paine, "Loading Probability Distributions in a Quantum circuit," arXiv:2208.13372v1 (2022).
\url{https://doi.org/10.48550/ARXIV.2208.13372}

\bibitem{Parl2021QSP}
Araujo, I.F., Park, D.K., Petruccione, F. et al., "A divide-and-conquer algorithm for quantum state preparation," Sci Rep, \textbf{11}, 6329 (2021).
\url{https://doi.org/10.1038/s41598-021-85474-1}

\bibitem{Zhang2022QSP}
Xiao-Ming Zhang, Tongyang Li, and Xiao Yuan, "Quantum State Preparation with Optimal Circuit Depth: Implementations and Applications," Phys. Rev. Lett., \textbf{129}, 230504 (2022).
\url{https://doi.org/10.1103/PhysRevLett.129.230504}

\bibitem{Yuan2023QSP}
Pei Yuan and Shengyu Zhang, "Optimal (controlled) quantum state preparation and improved unitary synthesis by quantum circuits with any number of ancillary qubits," Quantum, \textbf{7}, 956 (2023).
\url{https://doi.org/10.22331/q-2023-03-20-956}

\bibitem{Feynman:quantumcomputers}
R.P. Feynman, "Quantum mechanical computers," Found Phys, \textbf{16}, 507-531 (1986).
\url{https://doi.org/10.1007/BF01886518}

\bibitem{Aharonov1993}
Y. Aharonov, L. Davidovich, and N. Zagury, "Quantum random walks," Phys. Rev. A, \textbf{48}, 1687-1690 (1993).
\url{https://doi.org/10.1103/PhysRevA.48.1687}

\bibitem{Childs:quantumwalk}
Andrew M. Childs, "Universal Computation by Quantum Walk," Phys. Rev. Lett., \textbf{102}, 180501 (2009).
\url{https://doi.org/10.1103/PhysRevLett.102.180501}

\bibitem{Mallick:DCA2016}
A. Mallick, C. Chandrashekar, "Dirac Cellular Automaton from Split-step Quantum Walk," Sci Rep, \textbf{6}, 25779 (2016).
\url{https://doi.org/10.1038/srep25779}

\bibitem{Rajauria:DTQW}
Parth Rajauria, Prateek Chawla, and C. M. Chandrashekar, "Estimation of one-dimensional discrete-time quantum walk parameters by using machine learning algorithms," arXiv:2007.04572 (2020).
\url{https://doi.org/10.48550/ARXIV.2007.04572}



\bibitem{Venegas_Andraca_2012}
Salvador Elías Venegas-Andraca, "Quantum walks: a comprehensive review," Quantum Information Processing, \textbf{11}, 1015-1106 (2012).
\url{https://doi.org/10.1007/s11128-012-0432-5}

\bibitem{10.1145/780542.780552}
Andrew M. Childs, Richard Cleve, Enrico Deotto, et al., "Exponential Algorithmic Speedup by a Quantum Walk," Association for Computing Machinery, \textbf{10}, 59-68 (2003).
\url{https://doi.org/10.1145/780542.780552}

\bibitem{Childs2004}
Andrew M. Childs and Jeffrey Goldstone, "Spatial search by quantum walk," Phys. Rev. A, \textbf{70}, 022314 (2004).
\url{https://doi.org/10.1103/PhysRevA.70.022314}

\bibitem{Oxenfeldt1973}
A. R. Oxenfeldt, "A Decision-Making Structure for Price Decisions," Journal of Marketing, \textbf{37}(1), 48-51 (1973).
\url{https://doi.org/10.2307/1250774}

\bibitem{Ahmed2020}
Bouteska Ahmed, "Understanding the impact of investor sentiment on the price formation process: A review of the conduct of American stock markets," The Journal of Economic Asymmetries, \textbf{22}, e00172 (2020).
\url{https://doi.org/10.1016/j.jeca.2020.e00172}

\bibitem{Matsuzawa_2020}
Y. Matsuzawa, "An index theorem for split-step quantum walks," Quantum Information Processing, \textbf{19}, 227 (2020).
\url{https://doi.org/10.1007/s11128-020-02720-7}

\bibitem{Narimatsu:SSQW}
Akihiro Narimatsu, Hiromichi Ohno, and Kazuyuki Wada, "Unitary equivalence classes of split-step quantum walks," arXiv:2104.13529 (2023).
\url{https://doi.org/10.48550/ARXIV.2104.13529}

\bibitem{stroock_2010}
Daniel W. Stroock, "Probability theory: an analytic view," Cambridge University Press (2010).
\url{https://doi.org/10.1017/CBO9780511974243}

\bibitem{kallenberg2002foundations}
Olav Kallenberg, "Foundations of modern probability," Springer-Verlag, New York (2002).
\url{https://doi.org/10.1007/978-1-4757-4015-8}

\bibitem{BSmodel}
Fischer Black, Myron Scholes, "The Pricing of Options and Corporate Liabilities," Journal of Political Economy, \textbf{81}, 637–654 (1973).
\url{https://doi.org/10.1086/260062}

 
\end{thebibliography}
\end{document}